%% file: derandomizedLDT_arxiv.tex
\documentclass{article}

\usepackage{rpmacros}
\usepackage[dvipsnames,svgnames,x11names,hyperref]{xcolor}
\usepackage{amsthm}
\usepackage{xfrac}
\usepackage[T1]{fontenc}

\usepackage{gitinfo}
\usepackage{enumitem}
\usepackage{fullpage}
\usepackage{thmtools}
\usepackage{thm-restate}
\usepackage{nicefrac}
\usepackage[colorlinks,pagebackref]{hyperref}
\hypersetup{
  linkcolor=[rgb]{0,0,0.4},
  citecolor=[rgb]{0, 0.4, 0},
  urlcolor=[rgb]{0.6, 0, 0}
}
\usepackage{mathrsfs,bbm}
\usepackage{todonotes}
\usepackage{tikz}
\usetikzlibrary{decorations.pathreplacing,backgrounds}
\usepackage{multirow}
\usepackage{setspace}

\input{thmmacros}

\input{lazymacros}
\input{bibmacros}

\usepackage{thmtools}
\usepackage{thm-restate}

\newcommand*\samethanks[1][\value{footnote}]{\footnotemark[#1]}

\newif\ifanon
\anonfalse

\newif\ifdraft
\draftfalse

\ifdraft
\newcommand{\phnote}[1]{\textcolor{Brown}{\guillemotleft ph: #1 \guillemotright}}
\newcommand{\sriprahladhuvacha}[1]{\todo[color=red!100!green!33,inline,size=\small]{ph: #1}}
\newcommand{\RPnote}[1]{\textcolor{BrickRed}{\guillemotleft RP: #1 \guillemotright}}
\newcommand{\MKnote}[1]{\textcolor{Orange}{\guillemotleft MK: #1 \guillemotright}}
\newcommand{\MSnote}[1]{\textcolor{Blue}{\guillemotleft MS: #1 \guillemotright}}
\newcommand{\SKnote}[1]{\textcolor{Blue}{\guillemotleft SK: #1 \guillemotright}}
\newcommand{\gitinfonotecolour}{Gray}
\newcommand{\easteregg}{}
\else
\newcommand{\phnote}[1]{}
\newcommand{\sriprahladhuvacha}[1]{}
\newcommand{\RPnote}[1]{}
\newcommand{\MKnote}[1]{}
\newcommand{\MSnote}[1]{}
\newcommand{\SKnote}[1]{}
\newcommand{\gitinfonotecolour}{white}
\newcommand{\easteregg}{(Wht's th pnt f ths lne?)}
\fi

\newcommand{\ignore}[1]{}
\newcommand{\gitinfonote}{git info:~\gitAbbrevHash\;,\;(\gitAuthorIsoDate)\; \;\gitVtag}

\newcommand{\calL}{\mathscr{L}}
\newcommand{\calH}{\mathscr{H}}

\newcommand{\calP}{\mathscr{P}}
\newcommand{\calF}{\mathscr{F}}
\newcommand{\res}{\mathsf{Res}}

\DeclareMathOperator{\agr}{agree}
\let\agree\agr

\renewcommand{\epsilon}{\varepsilon}
\newcommand{\eps}{\epsilon}

\newcommand{\disc}{\mathsf{Disc}}
\newcommand{\HTr}{\mathsf{HTr}}
\newcommand{\Tr}{\mathsf{Tr}}

\newcommand{\calU}{\mathcal U}

\newcommand{\K}{\mathbb{K}}

\crefname{claim}{Claim}{Claims}

\title{On derandomizing the line-point low-degree test}
\author{Prahladh Harsha\thanks{Tata Institute of Fundamental Research, Mumbai, India. \href{mailto:prahladh@tifr.res.in}{\{prahladh,} \href{mailto:mrinal@tifr.res.in}{mrinal,} \href{mailto:ramprasad@tifr.res.in}{ramprasad\}@tifr.res.in}. Research supported by the Department of Atomic Energy, Government of India, under project number RTI400112, and in part by research grants from Google and Premji Invest.} 
\and 
Swastik Kopparty\thanks{Department of Computer Science and Department of  Mathematics, University of Toronto, Canada. \href{mailto:swastik.kopparty@utoronto.ca}{swastik@utoronto.ca}.}
\and 
Mrinal Kumar\samethanks[1]
\and 
Ramprasad Saptharishi\samethanks[1] 
\and 
Madhu Sudan\thanks{School of Engineering and Applied Sciences, Harvard University, Cambridge, Massachusetts, USA.\href{mailto:madhu@cs.harvard.edu}{madhu@cs.harvard.edu}. Supported in part by a Simons Investigator Award.}
}
\date{}

\begin{document}

\maketitle

\begin{abstract}
    We investigate the derandomized version of the classical line-point low-degree test in the high-error regime in the bivariate setting and reveal some unexpected behavior. 
    The classical line-point low-degree test takes an $m$-variate function $f$ over a finite field $\F_q$, queries it along a random line and accepts if the restriction of the function to this line is a low-degree polynomial.
    The high-error version of (the analysis of) this test asserts that if $f$ has at least $\epsilon$ expected local agreement with a degree-$k$ univariate polynomial on a random line, then in fact $f$ is non-trivially close to some degree-$k$ $m$-variate polynomial even as $\eps$ approaches $0$ (provided $q$ is sufficiently large compared to $k$ and $1/\eps$).  
    The derandomized version of the problem asks if the same holds when the line is not chosen uniformly over all possible lines but rather from a small set. Derandomized analyses of low-degree tests play a crucial role in obtaining near-linear sized PCPs. In this paper we explore the bivariate case ($m=2$).

    We show, somewhat surprisingly, that there exists a function $f\colon \F_q^2 \to \F_q$ and a set of lines $\calL$ in $\F_q^2$ of size nearly $q^{3/2}$ such that on every line in $\calL$, $f$ has agreement $\approx 1/2$ on average with some univariate polynomial of degree $k=\sqrt{q}/2$ and yet $f$ is $1 - O(q^{-1/2})$ far from every polynomial of degree $O(k)$. On the positive side, we show that once the field size $q$ is at least $k^2/\poly(\epsilon)$ such examples disappear and we can get sets of lines of size $O(q/\eps^{O(1)})$ on which the low-degree test has a high-error analysis. 

    On the way to our positive result, we show a new high-degree version of the classical two-direction Polishchuk--Spielman divisibility lemma. It deals with finding a low-degree global root $Z = U(X,Y)$ of a trivariate polynomial $Q(X,Y,Z)$, based on restrictions of $Q$ to lines in $r > \deg_Z(Q)$ different directions (generalizing the original $r = 2$ and $\deg_Z(Q) = 1$ case). 
    Specifically, it states that if $Q$ has a low-degree root $Z = U_{a,b}(X)$ for all lines $Y = aX + b$ in each of those $r$ directions, then $Q$ has a low-degree global root $Z = U(X,Y)$. This high degree Polishchuk--Spielman lemma is proved by a new technique: using simultaneous Hensel lifting of \emph{consistent} roots on many parallel lines. 

    Our negative result addresses one of the big challenges in analyzing low-degree tests, namely the rarity of interesting counterexamples that offer concrete barriers.
    
\end{abstract}

\newpage
\tableofcontents

\section{Introduction}\label{sec:intro}

In this paper, we study the problem of ``derandomizing'' the
``line-point'' low-degree test in the ``high-error'' regime. We show
a surprising limitation on how well this test can be derandomized over
fields of size quadratic in the degree parameter,
and complement this with an analysis of the derandomized line-point test
over fields of at least quadratic size (in the degree parameter). We expand on these terms and
results below.

Low-degree testing, one of the classical problems in property testing,
checks if a given function, viewed as a table of values, is close to the
evaluation of a polynomial of low total degree, by querying the function
only on a few points of the domain. More precisely, given a finite field
$\F$ of size $q$, a dimension parameter $m$ and a degree parameter $k$, the
goal of low-degree testing is to check if the function $f\colon \F^m \to \F$
is close to the evaluation of some $m$-variate polynomial $P$ of total
degree at most $k$, i.e., $P \in \calP_{k,m}$, where $\calP_{k,m}$ denotes the space
of all such polynomials, by querying the function $f$ on very few points of
$\F^m$.  The most natural such test is the \emph{line-point test} of
Rubinfeld and Sudan~\cite{RubinfeldS1996}: pick a uniformly random line
$\ell$ in $\F^m$, query $f$ at all points of $\ell$, and accept if the
restriction $f|_\ell$ is the evaluation of a univariate polynomial of degree
at most $k$ (i.e., $f|_\ell \in \calP_{k,1}$). Low-degree tests, and the
line-point test in particular, are the algebraic core of the proof of the
PCP theorem~\cite{AroraS1998, AroraLMSS1998}, and improvements to their
analyses have repeatedly translated into improvements in the parameters of
PCPs.

The soundness of the line-point test is typically studied in two regimes. In
the \emph{low-error} (or unique-decoding) regime, where the test rejects with
small constant probability $\delta$, the early works~\cite{RubinfeldS1996,
AroraS1998, AroraLMSS1998}, culminating in those of Polishchuk and
Spielman~\cite{PolishchukS1994} and Friedl and Sudan~\cite{FriedlS1995}, show
that $f$ is $O(\delta)$-close to a degree-$k$ polynomial as soon as
$q = \Omega(k)$. To state the results in the \emph{high-error} (or list-decoding) regime,
it is convenient to compare the expected local agreement of $f$ with
degree-$k$ univariate polynomials, namely
$\E_{\ell}[\agr(f|_\ell, \calP_{k,1})]$, with the global agreement
$\agr(f, \calP_{k,m})$.\footnote{Here, for functions $f, g$ on a common domain,
$\agr(f,g) := \Pr_{z}[f(z) = g(z)]$, where $z$ is uniform over the domain,
and for a family $\calF$ of functions,
$\agr(f,\calF) := \max_{g \in \calF} \agr(f,g)$.} For the line-point test over fields of size
$q \geq k^4/\poly(\epsilon)$, Arora and Sudan~\cite{AroraS2003} proved
that $\E_{\ell}[\agr(f|_\ell, \calP_{k,1})] \geq \epsilon$ implies
$\agr(f, \calP_{k,m}) \geq \poly(\epsilon)$. While this gave an analysis of
the line-point test in the high-error regime, it worked only when the field
size was at least quartic in the degree. In parallel and independent work,
Raz and Safra~\cite{RazS1997} proved a similar result, this time over
linear-sized fields (i.e., $q = \Omega(k)/\poly(\epsilon)$), at the
cost of replacing lines by planes (and hence $O(q)$ queries by $O(q^2)$
queries). In recent work~\cite{HarshaKSS2024}, a subset of the authors of this paper
gave a high-error analysis of the line-point test over fields of size
$\Omega(k)/\poly(\epsilon)$, thereby obtaining a common strengthening of
the above two results. In this work, we ask if this improved analysis of the
line-point test can be extended to a derandomized test.

\paragraph{Derandomized low-degree tests.}
A second, equally important, axis along which low-degree tests have been
improved is their \emph{randomness complexity}. The line-point test over
$\F^m$ picks a random line, which requires roughly $(2m-2)\log q$ random
bits, since there are about $q^{2m-2}$ lines in $\F^m$. When a low-degree
test is used inside a PCP, the randomness of the test essentially dictates
the size of the resulting proof, and so reducing the number of lines (or
planes) the test uses is the key step in constructing PCPs of nearly-linear
size~\cite{GoldreichS2006,PolishchukS1994, BenSassonSVW2003, BenSassonGHSV2006,BenSassonGHSV2005,MoshkovitzR2008,MoshkovitzR2010-linearpcp,MoshkovitzR2010-2qpcp}. In fact, it follows from Goldreich and Sudan~\cite{GoldreichS2006} that $|\F^m|^{1+o(1)}$ lines suffice (to be precise $|\F^m|\cdot \exp(\widetilde{O}(\sqrt{\log |\F^m|}))$ lines), however their proof does not yield an explicit set of such lines.
The
natural way to derandomize the test is to restrict the \emph{directions} of
the lines: instead of allowing every direction in $\F^m$, one allows only
directions from a small, structured set---for instance, directions from a
small-biased set~\cite{BenSassonSVW2003} or directions in $\mathbb{H}^m$ for
a subfield $\mathbb{H} \subseteq \F$~\cite{MoshkovitzR2008}. The latter
reduces the number of lines from about $q^{2m-2}$ to about $q^{m-1} \cdot
|\mathbb{H}|^{m-1}$.

In the low-error regime, derandomized line-point tests are well understood.
Already the Polishchuk--Spielman analysis~\cite{PolishchukS1994} is a
derandomized analysis: for bivariate functions it uses lines in only
\emph{two} directions (the axis-parallel lines), and it is this feature that
made it useful for nearly-linear sized PCPs. Ben-Sasson, Sudan, Vadhan and
Wigderson~\cite{BenSassonSVW2003} extended this to the $m$-variate
line-point test with directions drawn from a small-biased set. In the
high-error regime, however, the only derandomized low-degree test is the
derandomized \emph{plane-point} test of Moshkovitz and
Raz~\cite{MoshkovitzR2008}, which derandomizes the Raz--Safra
test~\cite{RazS1997} and which works over fields of size
$q = k/\poly(\epsilon)$ as long as the subfield satisfies
$|\mathbb{H}| \geq \poly(1/\epsilon)$. This leaves open the following
question.
\begin{quote}
  \emph{Is there a high-error analysis of a derandomized line-point test?}
\end{quote}

The Raz--Safra and Moshkovitz--Raz analyses are coding-theoretic and use the
fact that two planes typically intersect in a line, on which the restricted
polynomials form a code of distance $1 - o(1)$; two lines intersect in at
most one point, so this approach does not apply to line-point tests, which
instead have a very algebraic analysis~\cite{AroraS2003, HarshaKSS2024}.
One could, nevertheless, ask if the Moshkovitz--Raz approach to
derandomizing the plane-point test of Raz--Safra extends to the recent
improved analysis of the line-point test~\cite{HarshaKSS2024}. The
crucial observation of Moshkovitz--Raz is that besides the above
coding-theoretic fact, the Raz--Safra analysis uses just the sampling
properties of the planes-vs-points bipartite graph, which even the
derandomized-planes-vs-points graph has, and hence the Raz--Safra analysis
carries over. The algebraic analyses of the line-point
test~\cite{AroraS2003, HarshaKSS2024}, on the other hand, crucially use
lines in a large number of directions, a number that grows with the degree
$k$, whereas a derandomized test with $r$ directions offers only $r$ of
them, where $r$ should ideally depend only on $\epsilon$. This implies that
a Moshkovitz--Raz-style transformation of the analysis
of~\cite{HarshaKSS2024} is not feasible.

Nevertheless, in this work, we show that, with substantial modifications,
the analysis of~\cite{HarshaKSS2024} can be adapted to the setting where
$r$, the number of directions, is considerably smaller than the degree $k$.

\subsection{Our results}\label{sec:results-overview}

In this subsection, we describe our three results: (1) the necessity of quadratically large fields for derandomizing the line-point low-degree test, (2) an analysis of a
derandomized line-point test in which the number of directions satisfies
$r \ll k$ provided $q  \geq k^2/\poly(\epsilon)$, and (3) a generalization of the Polishchuk--Spielman divisibility lemma. We state all these results in the bivariate setting, i.e., $m = 2$
and $f\colon \F^2 \to \F$, since this case already presents several
challenges, as we will soon see. Unlike the analysis of the bivariate test in \cite{HarshaKSS2024}, we do not know if the analysis  in \autoref{thm:main technical} can be lifted from the bivariate to the multivariate case. 

\paragraph{Derandomization is infeasible in subquadratic-sized fields.}

The current best analyses of the line-point test~\cite{HarshaKSS2024} as well as the analysis of the derandomized plane-point test~\cite{MoshkovitzR2008} in the high-error regime work over fields of size
$k/\poly(\eps)$. However, a similar derandomized analysis of the line-point test has remained elusive. Our first result shows that such a derandomization using lines whose slopes come from a subfield is in fact infeasible as long as the field size is subquadratic in the degree parameter, at least for fields of small characteristic. This infeasibility result is especially interesting since counterexamples that offer concrete barriers are rare in the study of low-degree tests.

\begin{restatable}[Necessity of large fields for LDT]{theorem}{counterexampleLDT}\label{thm:intro-counterexample}
For every prime $p$, $\eps \in (0,1)$ of the form $p^{-r}$ and $s \in \N$, the following holds: Let $\F$ be the finite
field of size $q = p^{2rs}=(1/\eps)^{2s}$, $H$ be the subfield of $\F$ of
size $p^{rs}=(1/\eps)^{s}$ and $k = \eps\sqrt{q} = p^{r(s-1)}$ the degree parameter. For any $a, b \in \F$, let $\ell_{a,b}\colon \F \to \F^2$
denote the line given by $\ell_{a,b}(t) = (t, at + b)$. Let
$\calL = \{\ell_{a,b} \mid a \in H, b \in \F\}$. Observe that
$\calL$ consists of all lines with slope in the subfield $H$. 

There exist a function $f\colon \F^2 \to \F$ and univariate polynomials $P_\ell$ of
degree $k$, one for each $\ell \in \calL$, such that
$\E_{\ell \in \calL}[\agr(f|_\ell, P_\ell)] \geq \eps$ and yet
$\agr(f, \calP_{O(k),2}) \leq O(\nicefrac{\eps}{\sqrt{q}})$. 

Furthermore, there exists a trivariate polynomial $Q(X,Y,Z)$ of $Z$-degree
$1/\eps$ and $(1,1,k)$-degree $\sqrt{q}+1$ such that for each
$\ell \in \calL$, $P_\ell$ is a $Z$-root of the restriction of $Q$ to
$\ell$, yet $Q$ has no polynomial $Z$-root.
\end{restatable}

This choice of directions is precisely the subfield-based derandomization
of~\cite{MoshkovitzR2008}, thus ruling out a similar derandomization for the line-point test with comparable parameters.

The special case $\eps = \frac{1}{2}$ (i.e., $p=2$ and $r = 1$) shows that there exists a function $f$ having expected local agreement with degree $k$ univariate polynomials
as large as $\frac{1}{2}$, but still $f$ is $(1-o(1))$-far from bivariate polynomials of degree at most $O(k)$. 

An important caveat of these counterexamples is that the characteristic of the underlying field is small (in particular, $\text{char}(\F) \leq 1/\eps$), and it is a natural question to understand if these examples can be extended to finite fields of large characteristic. 

\paragraph{A high-error derandomized line-point test.}

We now complement the above counterexample by showing that once the field size is quadratically large, any set of $r$ directions for $r = \poly(1/\eps)$ suffices.

\begin{restatable}[Analysis of derandomized line-point test]{theorem}{LDTanalysismain}\label{thm:main technical}
There is a constant $c \in \N$ such that the following holds for every
$\epsilon > 0$, every finite field $\F$ of size $q$, and all natural numbers
$k, r$ satisfying $q \geq k^2/\epsilon^{c}$ and $r \geq 1/\epsilon^{c}$. Let
$U$ be a set of $r$ distinct directions in $\F^2$, and let $\calL$ be the set
of \emph{all} lines along these $r$ directions (i.e., $\abs{\calL} = rq$).
Then for any $f\colon \F^2 \to \F$, we have
\[
  \E_{\ell \in \calL}\bigl[\agr(f|_\ell, \calP_{k,1})\bigr] \geq \epsilon
  \implies \agr(f, \calP_{2k,2}) \geq \epsilon^{c}\,.
\]
\end{restatable}

Three features of this statement deserve comment.
\begin{itemize}
\item First, the number of directions must grow with $1/\epsilon$: if
  $g\colon \F \to \F$ is a random univariate function and
  $f(X,Y) := g(Y - mX)$, where the set of directions $U$ includes the
  direction with slope $m$, then $f$ is constant on every line with slope
  $m$, so the expected local agreement is at least $1/r$, and yet $f$ has
  negligible agreement with every low-degree polynomial. Thus
  $r > 1/\epsilon$ is necessary, and our theorem shows that
  $r = \poly(1/\epsilon)$ suffices.
\item Second, the conclusion shows agreement with a polynomial of degree
  $2k$ rather than $k$. Some loss of this kind is unavoidable once the
  directions are restricted: if the set $U$ consists of the directions given
  by $r$ distinct slopes $m_1, \ldots, m_r$ and $(r-1)$ divides $k$, the
  polynomial
  \[
    g(X,Y) := \prod_{i=1}^{r} (Y - m_i X)^{k/(r-1)}
  \]
  has degree at most $k$ on every line with slope in $U$, but has total
  degree $rk/(r-1) > k$. Thus $f = g$ passes the test with probability $1$, i.e., has
  perfect local agreement, but is far from every polynomial of degree $k$. In a companion paper, we
  show that the total degree of such polynomials is bounded above by
  $rk/(r-1)$. (The same phenomenon appears for $r = 2$ in the
  Polishchuk--Spielman analysis~\cite{PolishchukS1994}, which also concludes
  with a polynomial of degree $2k$.) 
\item Third, our requirement on the field size is quadratic in the degree
  $k$, unlike the derandomized plane-point test of
  Moshkovitz--Raz~\cite{MoshkovitzR2008}, which requires only a linear
  dependence. As observed in \cref{thm:intro-counterexample}, such a quadratic dependence is necessary at least for low-characteristic fields. More precisely, for a field $\F_q$ with characteristic at most $1/\eps$,  we require $q \geq (k/\eps)^2$ for derandomization to work with lines in even $\sqrt{q}$ directions, while the condition $q \geq k^2/\epsilon^{c}$ in the above theorem shows that this is tight up to the
  value of the constant $c$ and in this case, the number of directions can be made as low as $\poly(1/\eps)$. 
\end{itemize}

\paragraph{A Polishchuk--Spielman lemma for few directions and high $Z$-degree.}
At the heart of the Polishchuk--Spielman analysis~\cite{PolishchukS1994} is a
divisibility lemma, which can alternatively be stated
(see~\cite[Lemma~1.3]{HarshaKSS2024}) as follows. Let
$Q(X,Y,Z) = A(X,Y)Z - B(X,Y)$ be a trivariate polynomial with $Z$-degree
one. Suppose the restriction of $Q$ to many axis-parallel lines has a
$Z$-root that is a polynomial of degree at most $k$; in other words, for
various restrictions of the form $X = a$ or $Y = b$, the restricted
polynomial $A$ divides the restricted polynomial $B$ and the quotient has
degree at most $k$. Then $Q$ itself has a $Z$-root that is a bivariate
polynomial of degree at most $k$ in each variable; that is, $A$ divides $B$
and the quotient has degree at most $k$ in each variable. Intuitively, the global root is what explains the
roots on the various restrictions.

A natural question is whether this lemma generalizes to polynomials $Q$ of
$Z$-degree larger than one. Such a generalization was crucial to the
analysis of the line-point test in the high-error regime
in~\cite{HarshaKSS2024} (see~\cite[Lemma~1.4]{HarshaKSS2024}). However, the
generalization there required, as a hypothesis, that $Q$ have a low-degree
$Z$-root on its restrictions to lines in many directions (not just two, as
in the original Polishchuk--Spielman setting). One could ask for the
smallest number of directions for which such a generalization holds. If
the number of directions equals the $Z$-degree, i.e., $r = \deg_Z(Q)$, then
the statement is simply false: for distinct $m_1, \ldots, m_r$ and
irreducible $g_1, \ldots, g_r$ of degree $k$, the polynomial
\[
  Q(X,Y,Z) := \prod_{i=1}^{r} \bigl((Y - m_iX)Z - g_i(X,Y)\bigr)
\]
has a degree-$k$ $Z$-root on every line $Y = m_iX + c$ with $c \neq 0$, but
no polynomial $Z$-root. (The case $r = 2$ of this example appears
in~\cite{HarshaKSS2024}.) 

Our third result shows that this example is essentially the only
obstruction: as soon as the number of directions exceeds the $Z$-degree,
the generalized lemma holds, provided the field is large enough.

\begin{lemma}[Generalized Polishchuk--Spielman lemma]\label{thm:PS-gen-intro}
Let $Q(X,Y,Z) \in \F[X,Y,Z]$ have $(1,1,k)$-degree $D$, and
suppose that the restriction of $Q$ to every line in $r > \deg_Z(Q)$
directions has a $Z$-root that is a univariate polynomial of degree at most
$k$. If $q \geq \poly(\deg_Z(Q)) \cdot Dk$, then $Q$ has a $Z$-root
$P(X,Y) \in \F[X,Y]$ of total degree at most $2k$. Moreover, there exist two
directions (among the $r$ directions) such that the restriction of $P$
agrees with the given roots on all but $O(Dk)$ of the lines in these two
directions.
\end{lemma}

Compared to the corresponding statement in~\cite[Lemma~1.4]{HarshaKSS2024},
this result needs only $\deg_Z(Q) + 1$ directions, which is optimal by the
example above, rather than a number of directions that grows with $D$. This
comes at a cost, however: the field must be of size $\Omega(Dk)$ rather
than $\Omega(D + k)$, and, as observed in \cref{thm:intro-counterexample}, this large field size is
unavoidable and the condition
$q \geq \poly(\deg_Z(Q)) \cdot Dk$ cannot be weakened to $q = O(D + k)$.

Our proof of \cref{thm:main technical} does not directly use this lemma.
Instead, both statements are proved from another, similar statement
(see \cref{lem:lifting-parallel-lines-general}), which shows that consistent low-degree $Z$-roots
on a family of parallel lines can be lifted to a global \emph{rational}
$Z$-root of $Q$. \cref{thm:PS-gen-intro} follows from \cref{lem:lifting-parallel-lines-general} by 
observing that if there are $\deg_Z(Q) + 1$ sets of parallel lines, then there is a pair of directions such that the roots on the lines in these directions agree on a constant (depending on $\deg_Z(Q)$) fraction of points in $\F^2$. This agreement provides the necessary structure for   us to apply \cref{lem:lifting-parallel-lines-general} for lines in these two directions. This also allows us to conclude that the two rational roots obtained must in-fact be the same! We then use the standard Polischuk-Spielman lemma to conclude that these rational roots must be polynomials. 

Given that our counterexamples only prove tightness of the above result when the characteristic is small (in particular when $\text{char}(\F) \leq 1/\eps$), it would be interesting to know whether the quantitative bounds in \autoref{thm:main technical} and \autoref{thm:PS-gen-intro} can be improved over fields of sufficiently large characteristic.

\subsection*{Organization}
In the next section, we provide an overview of our proofs. We set up the necessary notation and preliminaries in \autoref{sec:prelims}. We then prove \autoref{lem:lifting-parallel-lines-general} in \autoref{sec:lifting-lemma}. We use this lemma to complete the proof of \autoref{thm:PS-gen-intro} in \autoref{sec:PS-gen} and the analysis of the derandomized low-degree test in \autoref{thm:main technical} in \autoref{sec:LDT-analysis}. Finally, we prove \autoref{thm:intro-counterexample} in \autoref{sec:counterexamples}. 

\subsection*{AI use}
All the technical content in the paper is due to its human authors. An AI assistant (Claude) was used to help write the introduction, and for typographical corrections and minor edits to the abstract and the proof overview.

\section{Overview of the proofs}\label{sec:proof-overview}

Our proofs of \autoref{thm:main technical} and \cref{thm:PS-gen-intro} follow the high-level outlines of similar results in \cite{AroraS2003} and \cite{HarshaKSS2024}. However, the presence of lines in very few directions introduces some important technical challenges that the proof has to overcome. In this section, we discuss the high-level outline of the proofs of the main theorems. 

\paragraph{Proof of \autoref{thm:main technical}: }
Let $f\colon \F^2 \to \F$ be a function that satisfies
\[ \E_{\ell \in \calL}[\agr(f|_\ell,\calP_{k,1})]\geq \eps\,,\]
where $\calL$ refers to the set of all lines in $r$ distinct directions. The proof starts by focusing on the set of lines in $\calL$ and points in $\F^2$ that pass the test with probability $\Omega(\epsilon)$. We call such points and lines \emph{good} and focus on them for the analysis of the test. We show that there is a non-zero trivariate $Q(X,Y,Z)$ of \emph{low} total degree and \emph{constant} $Z$-degree such that for \emph{all} good points $\va \in \F^2$, $Q(\va, f(\va))$ is zero. We also show that for all good lines $\ell$, there is a univariate $g_\ell$ of degree $k$ such that $g_\ell$ is a $Z$-root of the polynomial $Q((X,Y)_\ell,Z)$. By massaging $Q$ to avoid some degenerate cases, we show that $Q$ satisfies the hypothesis of a variant of a generalization (\autoref{lem:lifting-parallel-lines-general}) of the Polishchuk--Spielman lemma. We then rely on this lemma, together with bounds on the degree of $Q$ to complete the proof of \autoref{thm:main technical}. The details can be found in \autoref{sec:LDT-analysis}. 

We note that while the interpolation step is similar to that in \cite{AroraS2003,HarshaKSS2024}, the details turn out to be substantially cleaner in the proof here. Instead of focussing on two good directions as is done in the earlier proofs, we start by sampling a small number of good lines $\ell$ and finding a non-zero low-degree $Q$ such that $Q((X,Y)_\ell,g_\ell)$ is identically zero for these special lines. We then use the geometry of the space $\F^2$ to conclude that $Q$ must satisfy a similar constraint for \emph{all} good lines $\ell$. A similar interpolation argument was also independently discovered in the recent work of Kominers, Thaler and Zheng \cite{KominersTZ2026} on an improved analysis of the (non-derandomized) line-point test. 

The main technical highlight of this proof is a version of the Polishchuk--Spielman lemma in the high $Z$-degree setting. We now describe the lemma, the main ideas in its proof and the proof of \autoref{thm:PS-gen-intro} which improves on the corresponding lemma in \cite{HarshaKSS2024} in the number of directions required: only one more than the $Z$-degree, which is the bare minimum.

\paragraph{Proof of \autoref{thm:PS-gen-intro}: }
The following technical lemma is the key to both the proof of \autoref{thm:main technical} and \autoref{thm:PS-gen-intro}. 

\begin{restatable}[Lifting from consistent parallel lines]{lemma}{consistentparallellineslift}\label{lem:lifting-parallel-lines-general}
    There is a constant $c> 0$ such that the following holds. Let $Q(X,Y,Z)$ be a non-zero trivariate polynomial of $(1,1,k)$-degree $D$. Let $\calL$ be a set of parallel lines in $\F^2$ such that for every $L \in \calL$, there exists a polynomial $g_L$ of degree at most $k$ that satisfies 
 \[
 Q((X,Y)_L,g_L) \equiv 0 . 
 \]
 Furthermore, let $\hat{L}$ be a line that is not parallel to the lines in $\calL$, and let $\hat{h}$ be a degree $k$ polynomial such that $Q((X,Y)_{\hat{L}},\hat{h})$ is identically zero, $(\partial_Z Q)((X,Y)_{\hat{L}},\hat{h})$ is non-zero, and for every $L \in \calL$, $\hat{h}$ and $g_L$ agree at the point $L\cap \hat{L}$. 

 If $|\calL| > c \cdot (\deg_Z(Q)Dk)$, then there exists $U(X,Y) \in \F(X, Y)$ of total degree at most $Dk$ such that
 \[
 Q(X,Y,U(X,Y)) \equiv 0. 
 \] 
 Moreover, for all but at most $D$ choices of $L \in \calL$, the restriction of $U$ on $L$ is well defined and equals $g_L$.     
 
 \end{restatable}
This lemma shows that consistent, low-degree $Z$-roots of $Q(X,Y,Z)$ on a set of parallel lines can be lifted to a common global rational root. Here, consistency of the roots on the parallel lines is asserted by their agreement with the $Z$-root on a separate anchor line $\hat{L}$ that intersects these parallel lines, and the fact that the multiplicity of the root on $\hat{L}$ is one. A similar lifting theorem, but from a non-degenerate point, is at the heart of the proof in \cite{HarshaKSS2024}. We need to lift from an anchor line instead of an anchor point  as in \cite{HarshaKSS2024} as the number of lines passing through any point in the derandomized test is very small (in fact, $\ll k$). 
Lifting from an anchor point $(\alpha,\beta)$ involves division by $(\partial_Z Q)(\alpha,\beta,f(\alpha,\beta))$ and hence the only requirement is that this field element is non-zero. Lifting from an anchor line $\hat{L}$ involves division by the polynomial $(\partial_Z Q)((X,Y)_{\hat{L}},\hat{h})$ and hence the requirement that this polynomial is non-zero. However, this causes several complications as we are dividing by a polynomial and not a field element. We need to ensure that the degrees of the numerator and denominator polynomials do not blow up and one has to carefully count the degree growth. This is done using a corollary of a result of Furstenberg~\cite{Furstenberg1967,BhattacharjeeKRRSS2026} applied to Newton's iteration (see \cref{thm:NI-explicit}).

To prove the above lemma, we show that the root on the anchor line can be lifted via Newton iteration to obtain a global power series root. This power series root is then shown to be equal to the low-degree polynomial roots on (most of) the lines in $\calL$ using the consistency of the anchor line with the lines in $\calL$. We then work with an appropriate truncation of the power series root (which has rational coefficients) to show that if the number of lines in $\calL$ is large enough, then this truncation gives us a rational global $Z$-root of $Q$. The details of the proof are discussed in \autoref{sec:lifting-lemma}. 

To prove \autoref{thm:PS-gen-intro}, we start by preprocessing $Q$ to ensure some non-degeneracy properties (e.g. square-freeness and non-zero $Z$-derivative). We then use the fact that the $Z$-degree of $Q$ is strictly smaller than the number of directions available to conclude that there is a pair of directions (among the $r$ available directions), such that roots on many lines in these directions are consistent (i.e. their values on the intersection points agree) for at least $\Omega(q^2/\poly(\deg_Z(Q)))$ of the points in $\F^2$. For simplicity of notation, let us assume that the directions correspond to lines parallel to the $X$ and $Y$-axes (i.e, horizontal and vertical lines). The high-level idea is to invoke \autoref{lem:lifting-parallel-lines-general} twice: once on a set of horizontal parallel lines with a vertical line as anchor, and once on a set of vertical parallel lines with a horizontal line as anchor.
This gives us two global roots $U(X,Y), U'(X,Y)$ of $Q(X,Y,Z)$: $U(X,Y)$ lies in $\F(Y)[X]$  with $X$-degree at most $k$, and $U'(X,Y)$ lies in $\F(X)[Y]$ with $Y$-degree at most $k$.
Moreover, these roots {\em explain} the given roots of $Q$ on many horizontal lines and vertical lines respectively.
Since we chose the two directions to be such that the given roots of $Q$ on lines in these directions are often consistent, this implies that $U(X,Y)$ and $U'(X,Y)$ agree on many points. By the Schwartz-Zippel lemma, this means that the low-degree rational functions $U(X,Y)$ and $U'(X,Y)$ are identical, and in particular they lie in $\F(Y)[X] \cap \F(X)[Y] = \F[X,Y]$ and have degree at most $k$ in both variables.

Filling in the details, e.g. handling degeneracy of $Q$, showing the existence of good anchors for the lifting outlined above and dealing with lines where the rational root does not equal the polynomial root on the line requires some care, and we discuss the details of the proof in \autoref{sec:PS-gen}.

\paragraph{Proof of \autoref{thm:intro-counterexample}}
For the proof of \autoref{thm:intro-counterexample}, let us for the purpose of this overview restrict our attention to characteristic two fields (i.e., $p=2$) and when $\eps = \nicefrac12$ (i.e., $r=1$). In this case, we work over a field $\F_{q}$, where $q = 2^{2s}$. This field clearly contains the subfield $\F_{2^{s}}$ of size $\sqrt{q}$. We set the degree parameter $k$ to $\sqrt{q}/2$. The lines considered for the line-point test in this construction are of the form $\ell_{a,b} = \{(t, at + b): t \in \F_q\}$, where $a$ is in $\F_{\sqrt{q}}$ and $b$ is in $\F_q$. Thus, the set $\calL$ of lines considered here has $q^{3/2}$ lines in it, $q$ lines each in the $|\F_{2^s}|=\sqrt{q}$ directions. To construct the function $f\colon \F_{q}^2 \to \F_q$, we start by assigning univariate polynomials of degree $k$ to each of the lines in $\calL$. We show that over a uniformly random pair of lines chosen from $\calL$, the line polynomials agree at the intersection of these lines with probability very close to half, and yet there is no degree $O(k)$ bivariate polynomial whose restriction to a constant fraction of the lines $\ell$ in $\calL$ equals the line polynomial in $\ell$. These properties are very closely built into the careful construction of the line polynomials, which are constructed using properties of the trace function ($\Tr$) from $\F_q$ to $\F_2$. More precisely, the line polynomial associated to the line $\ell_{a,b}$ for $a \in \F_{\sqrt{q}}$ and $b \in \F_q$, is the polynomial $\HTr(b^{\sqrt{q}}T)$, where the \emph{half-trace} polynomial $\HTr(Z)$ is defined as 
\[
\HTr(Z) = Z + Z^2 + Z^4+ \dots + Z^{2^{s-1}} . 
\]
Observe that $\HTr$ is the trace map $\Tr_{\F_{2^s}/\F_2}$ extended as a polynomial map to all of $\F_q = \F_{2^{2s}}$ and is related to $\Tr$ as $\HTr(z) + \HTr(z^{\sqrt{q}}) = \Tr(z)$ for every $z \in \F_q = \F_{2^{2s}}$. 

On the way to the proof of \autoref{thm:intro-counterexample}, we also construct a polynomial $Q(X,Y,Z)$ defined as 
\[
Q(X,Y,Z) := Z^2 + Z + (XY^{\sqrt{q}} + YX^{\sqrt{q}}),  
\]
with $Z$-degree $2$ and $(1,1,k)$-degree $\sqrt{q}+1$. We show that on every line $\ell_{a,b}\in \calL$, the polynomial $\HTr(b^{\sqrt{q}}T)$ is a $Z$-root of $Q((X,Y)_{\ell_{a,b}},Z)$, and yet $Q(X,Y,Z)$ has no $Z$-root in the bivariate polynomial ring $\F[X,Y]$. 

Note that we are considering lines in $\sqrt{q}$ directions here, and yet the $Z$-degree of $Q$ is just $2$! This is in stark contrast to the setting of \autoref{thm:PS-gen-intro}, where over a field of sufficiently large (and yet quadratic) size, existence of low-degree polynomial $Z$-roots on all lines in any set of $(\deg_Z(Q)+1)$ directions is sufficient to guarantee the existence of a low-degree polynomial $Z$-root globally.

This overview focuses on the case when $p=2$ and $r=1$ (and hence $\eps = 1/2$). Our results work for all characteristics and all $\eps$ provided the characteristic is at most $1/\eps$. We discuss the details of these constructions in \autoref{sec:counterexamples}.

\section{Notations and preliminaries}\label{sec:prelims}

\subsection{Notations}
\begin{itemize}
    \item $\F$ denotes a (usually) finite field. 
    \item The support of a polynomial is the set of non-zero monomials in it. 
    \item $\F[X]_{<d}$ denotes the linear space of univariate polynomials of degree less than $d$ with coefficients in $\F$, and can naturally be identified with the space $\F^d$. 
    \item The $(1,1,k)$-weighted degree (or simply the $(1,1,k)$-degree) of a monomial $X^iY^jZ^{\ell}$ is defined to be equal to $(i + j + k\ell)$. The $(1,1,k)$-degree of a polynomial $Q(X,Y,Z)$ is defined to be the maximum of the $(1,1,k)$-degrees of any monomial in the support of $Q$. 
    \item The degree of a rational function $R(X)$ is defined to be equal to the maximum of the degrees of the numerator of $R$ and the denominator of $R$.
    \item For $\ell \in \mathbb{N}$ and polynomial $Q(Y,Z) \in \K[Y,Z]$, the expression $[Z^{\ell}]\{Q(Y,Z)\}$ denotes the coefficient $Z^{\ell}$ in the polynomial $Q(Y,Z)$, when it is viewed as a univariate in $Z$ with coefficients in $\F[Y]$. 
    \item For a trivariate polynomial $Q(X,Y,Z)$ and a line $L \in \F^2$, the polynomial $Q((X,Y)_L,Z)$ denotes a bivariate polynomial obtained by restricting the variables $(X,Y)$ to the line $L$. For instance, if the line $L$ is parameterized as $(T, mT+c)$ for slope $m$ and intercept $c$, $Q((X,Y)_L,Z)$ denotes the bivariate polynomial $Q(T,mT+c, Z)$. 
    \item For a line $L$ and a trivariate $Q(X,Y,Z)$, we say that $Q$ has a $Z$-root of degree $k$ on the line $L$ if there is a polynomial $h(T)$ of degree at most $k$ such that $Q(T,mT+c,h(T))$ is identically zero, where $L = \{(t, mt+c): t \in \F\}$.  

    We note that the existence of a degree $k$ $Z$-root on a line $L$ is independent of the precise parameterization of the line $L$ and is a property of the polynomial $Q$. Thus, we often just say that a polynomial $h$ is a $Z$-root of $Q$ on the line $L$, indicated by $Q((X,Y)_L,h) \equiv 0$, without specifying the precise parameterization.

    \end{itemize}

\subsection{Preliminaries}

\begin{lemma}[Chernoff bound]\label{lem:chernoff}
    Let $X_1, X_2, \dots, X_n$ be independent random variables taking values in $\{0,1\}$,  let $X$ denote their sum, and $\mu$ be the expected value of $X$. Then, \[
    \Pr\left[ |X - \mu| \geq \delta \mu\right] \leq 2\cdot \exp(-\delta^2 \mu/3),  
    \] 
    for every $\delta \in [0,1]$. 
\end{lemma}

\begin{lemma}[Claims A.4, A.5, A.6 in \cite{HarshaKSS2024}]\label{lem:monomial-count}
    For all $p, k, D \in \N$ such that $D > 20k$ and $p \geq 2$. 
    
    Then, the number of monomials $X^{i}Y^jZ^{\ell}$ of $(1,1,k)$-degree at most $D$ is greater than $D^3/6k$, and the number of monomials $X^{i}Y^jZ^{\ell}$ of $(1,1,k)$-degree at most $D$, where $\ell$ is either zero or is not divisible by $p$ is greater than $D^3/12k$.

\end{lemma}

\begin{lemma}[Schwartz-Zippel lemma]\label{lem:SZ}
    Let $\F$ be any field and $A$ be a subset of $\F$. If $P(\vx)$ is a non-zero polynomial on $n$ variables with coefficients in $\F$, then the number of zeroes of $P$ on the product set $A^n$ is at most $\left(\deg(P)\cdot |A|^n\right)$. 
\end{lemma}

\begin{lemma}\label{lem:linear-factors}
Let $\F$ be any field and $f(X,Y) \in \F[X,Y]$ be a non-zero polynomial of total degree at most $D$. Then, the number of lines $L$ in $\F^2$ such that the restriction of $f(X,Y)$ on $L$ is identically zero is at most $D$.  
\end{lemma}
\begin{proof}
    Each distinct line on which $f$ restricts to $0$ gives a distinct linear factor of $f(X,Y)$.
\end{proof}

\begin{definition}\label{def:square-free-part}
    Let $f \in \F[\mathbf{X}]$ be a polynomial that factorizes as $f = \prod_{i} f_i(\mathbf{X})^{e_i}$, for irreducible polynomials $f_i$ such that $\gcd(f_i,f_j) = 1$ for $i \neq j$, and natural numbers $e_i \geq 1$. Then, the squarefree part of $f$ is defined as $\prod_i f_i$.  
\end{definition}

\subsection{Hasse derivatives}
\begin{definition}
    The Hasse derivative of order $t$ of a polynomial $f(X) \in \F[X]$, denoted by $\partial_{X^t}(f)$, is defined as the coefficient of $Z^t$ in the polynomial $f(X+Z)$, when $f(X+Z)$ is viewed as a univariate in $Z$ with coefficients in the ring $\F[X]$. 
\end{definition}

\begin{lemma}[Properties of Hasse derivatives]\label{lem:hasse-props}
    Let $\F$ be a field of characteristic $p$. Then, the following properties hold. 
    \begin{itemize}
        \item    Let $f(X) \in \F[X]$ be a non-zero polynomial of degree at least one. The first order Hasse derivative of $f$ is identically zero, if and only if there exists a polynomial $g(X) \in \F[X]$ such that $g(X^p) = f(X)$. 
        \item For any two polynomials $f, g$, $\partial_X(fg) = g \cdot \partial_X(f) + f\cdot \partial_X(g)$. By induction, this implies that for every $n \in \N$, $\partial_X(f^n) = nf^{n-1}\partial_X(f)$. 
    \end{itemize}
 
\end{lemma}

A consequence of the above lemma is the following corollary. 

\begin{corollary}\label{obs:nonzero-derivative-for-roots}
Let $\F$ be any field, $n \in \N$ be a natural number and $f, g \in \F[Z]$ be non-zero polynomials such that $f = g^n$. 

If $n$ is not divisible by the characteristic of the field and $\partial_Z(f)$ is non-zero, then $\partial_Z(g)$ is non-zero. 
\end{corollary}
\begin{proof}
    This immediately follows from the fact that $\partial_Z(f) = ng^{n-1}\partial_Z(g)$ in \autoref{lem:hasse-props}. 
\end{proof}

In the next lemma, we try to understand the following issue: we have a bivariate $f(X,Y) \in \F[X,Y]$ over a field of characteristic $p$ that is \emph{not} the $p$th power of a polynomial. The goal of the lemma is to understand the number of lines $L$ such that the restriction of $f$ on the line $L$ is the $p$th power of a polynomial. 
\begin{lemma}\label{lem:pth-power-on-lines}
Let $\F$ be any field of characteristic $p$ and $f(X,Y) \in \F[X,Y]$ be a non-zero polynomial of total degree at most $D$ such that there \emph{does not exist} a polynomial $g(X,Y) \in \F[X,Y]$ satisfying $f(X,Y) = g(X,Y)^p$. 

Then, there is at most one direction $\Delta \in \F^2$ such that for all directions $\Delta' \neq \Delta$, the following is true: the number of lines $L$ in the direction $\Delta'$ for which the restriction of $f$ on $L$ is $p$th power of a polynomial is at most $D$. 
\end{lemma}
\begin{proof}
    Let $f = \sum_{i, j} f_{i,j}X^iY^j$ for non-negative integers $i,j$ such that $i+j \leq D$, and $f_{i,j} \in \F$. Since $f$ is not the $p$th power of a polynomial, there exists an $f_{i,j}$ that is non-zero, such that $p$ does not divide at least one of $i$ and $j$. Thus, from \autoref{lem:hasse-props}, we get that at least one of the first order Hasse derivatives $\partial_X(f)$ or $\partial_Y(f)$ is not identically zero. Without loss of generality, we assume that $\partial_Y(f)$ is non-zero. 

    \paragraph*{Lines parallel to $Y$-axis: } Since $\partial_Y(f)$ is non-zero, there is at least one non-zero monomial $f_{i_0,j_0}$ in $f$ such that $p$ does not divide $j_0$. Let $f_{j_0}(X)$ be the coefficient of $Y^{j_0}$ in $f$, when it is viewed as a univariate in $Y$ with coefficients in $\F[X]$. Clearly, for an $a \in \F$, $f(a,Y)$ is the $p$th power of a univariate in $\F[Y]$ only if $f_{j_0}(a)$ is zero. But there are at most $\deg_X(f) \leq D$ many such $a$.  

    \paragraph*{Lines of the form $(Y = mX+c)$ for $m \in \F$: } Since $\partial_Y(f)$ is non-zero, there is at most one $m_0 \in \F$ such that the polynomial $\partial_X(f)(X,Y) + m_0\partial_Y(f)(X,Y)$ is identically zero. We now argue that for every $m \neq m_0$, the set of lines $L$ of slope $m$ such that the restriction $f_L$ is a $p$th power has size at most $D$. 

    From \autoref{lem:hasse-props} and the definition of Hasse derivatives, we get that 
    \[
    \partial_X(f(X,mX+c)) = \partial_X(f)(X,mX+c) + m\partial_Y(f)(X,mX+c) \, .
    \]
    Furthermore, for $f_L = f(X,mX+c)$ to be a $p$th power, it must be the case that the Hasse derivative $\partial_X(f(X,mX+c))$ is identically zero, or in other words, $c$ is a zero of the polynomial $\partial_X(f)(X,mX+Z) + m\partial_Y(f)(X,mX+Z) \in (\F[X])[Z]$. But this polynomial has $Z$-degree at most $D$, and is non-zero; indeed $m \neq m_0$ ensures that $\partial_X(f)(X,Z) + m\partial_Y(f)(X,Z)$ is non-zero, and shifting $Z$ to $Z + mX$ retains this non-zeroness. Thus, there at most $D$ choices of $c$ for which $f_L$ is a $p$th power. 
\end{proof}

\subsection{Resultant and discriminant}
\begin{definition}[Sylvester Matrix, Resultant and Discriminant]
    The Sylvester matrix of two univariates $f(X), g(X)$ over a field $\F$ of degree $d_1 \geq 1, d_2 \geq 1$ is a $(d_1+d_2)\times (d_1+d_2)$ matrix corresponding to the linear transformation  
    \[
    \Gamma_{f,g}:\F^{d_2} \times \F^{d_1} \to \F^{d_1+d_2}, 
    \]
    defined as 
    \[
    \Gamma_{f,g}(A,B) = Af + Bg \, ,
    \]
    where $A, B \in \F[X]$ are polynomials of degree at most $d_2-1$ and $d_1-1$ respectively. 

    The determinant of the Sylvester matrix of $f,g$ is referred to as the Resultant of $f,g$ and is denoted by $\res(f,g)$. The resultant of $f$ and its first order Hasse derivative $f^{(1)}$ is referred to as the Discriminant of $f$ and is denoted by $\disc(f)$. 
\end{definition}
The utility of the resultant and discriminant stem from the following fundamental facts. 
\begin{theorem}\label{thm:res-and-gcd}
For every field $\F$ and polynomials $f,g \in \F[X]$ of degree at least one, the $\res(f,g)$ is non-zero if and only if $f,g$ have $\gcd$ $1$. 
\end{theorem}

\begin{corollary}\label{thm:disc-and-sqr-free}
Let $\F$ be a finite field or a field of characteristic zero and $f \in \F[X]$ be a polynomial of degree at least two. Then, $f$ is squarefree if and only if $\disc(f)$ is non-zero.  
\end{corollary}
In the course of our proofs, we would also have to work with multivariates that we view as univariates in one of the variables with coefficients in the field of rational functions. For such fields, we have the following weaker variant of \autoref{thm:disc-and-sqr-free}.  

\begin{corollary}\label{thm:disc-and-sqr-free-multivariates}
Let $\F$ be any field and $f \in \F[X]$ be a polynomial of degree at least two such that the Hasse derivative $\partial(f)$ is non-zero and has degree at least one. Then, $f$ is squarefree if and only if $\disc(f)$ is non-zero.  
\end{corollary}

On many occasions in this paper, we work with polynomials with larger number of variables and view them as univariates with respect to one of the variables and coefficients coming from the field of rational functions in the other variables. In such settings, we use $\res_Z(f,g)$ and $\disc_Z(f)$ to denote the resultant and discriminants respectively, with the subscript indicating that we are viewing the multivariates as univariates in $Z$. 

Another simple, but useful structural property of the resultant (and discriminant) for multivariate polynomials that we rely on is the following. 

\begin{observation}\label{obs:resultant-restrictions-commute}
Let $f(X,Y, Z), g(X,Y,Z) \in \F[X,Y,Z]$ be trivariates with $Z$-degree at least one, and let $\va \in \F^2$ be such that $\deg_Z(f(X,Y,Z)) = \deg_Z(f(\va,Z))$ and $\deg_Z(g(X,Y,Z)) = \deg_Z(g(\va,Z))$. 

Then, $\res_Z(f(X,Y,Z),g(X,Y,Z)) \in \F[X,Y]$ when evaluated at $\va$ equals the $\res_Z(f(\va,Z),g(\va,Z))$. 
\end{observation}
\begin{proof}
    The proof follows from the fact that the $Z$-degree preserving property of $\va$ ensures that the Sylvester matrix of $f(X,Y,Z), g(X,Y,Z)$ (while viewing these as a univariate in $Z$) when evaluated at $\va$ equals the Sylvester matrix of $f(\va,Z), g(\va,Z)$. 
\end{proof}

\subsection{Power series roots}

\begin{theorem}[Newton iteration]\label{thm:NI}
Let $\F$ be any field and $\K$ be any extension of $\F$, and let $P(Y,Z) \in \F[Y,Z]$ be a bivariate polynomial with coefficients in $\F$. Let $a \in \K$  be such that $P(0,a) = 0$ and $\partial_Z(P)(0,a) \neq 0$. Let $\Phi_0 := a$, and for every $k \geq 1$, $\Phi_{k}(Y) \in \K[Y]$ be defined as 
\[
\Phi_k(Y) := \left( \Phi_{k-1}(Y) - \frac{P(Y, \Phi_{k-1}(Y))}{\partial_Z(P)(0,a)} \right) \bmod Y^{k+1}. 
\]
Then, for every $k \geq 1$, $\Phi_k(Y)$ is the \emph{unique} polynomial that satisfies the following conditions.  
\begin{align*}
&P(Y,\Phi_k(Y)) \equiv 0 \bmod Y^{k+1},    \\
&\Phi(Y) \equiv a \bmod Y, \quad \quad \text{ and } \\
&\deg(\Phi_k(Y)) \leq k. 
\end{align*}
\end{theorem}
The following theorem gives an explicit non-iterative expression for the approximate root $\Phi_k(Y)$ defined above. This is a corollary of a theorem of Furstenberg \cite{Furstenberg1967} and the precise expression here is implicit in Corollary 3.3 and Theorem 4.1 in \cite{BhattacharjeeKRRSS2026}.  
\begin{theorem}[Explicit Newton iteration \cite{Furstenberg1967,BhattacharjeeKRRSS2026}]\label{thm:NI-explicit}
    Let $\F$ be any field and $\K$ be any extension of $\F$, and let $P(Y,Z) \in \F[Y,Z]$ be a bivariate polynomial with coefficients in $\F$. Let $a \in \K$  be such that $P(0,a) = 0$ and $\partial_Z(P)(0,a) \neq 0$. Let $\Phi_0 := a$, and for every $k \geq 1$, $\Phi_{k}(Y) \in \K[Y]$ be defined as 
\[
\Phi_k(Y) := \left( \Phi_{k-1}(Y) - \frac{P(Y, \Phi_{k-1}(Y))}{\partial_Z(P)(0,a)} \right) \bmod Y^{k+1}. 
\]
Then, $\Phi_k(Y)$ equals the polynomial
\[
\left(\sum_{m = 1}^{2k} \frac{1}{\alpha^{m+1}}[Z^{m-1}] \{ \partial_Z(P(Y,Z+a))\dot (\alpha Z - P(Y,Z+a))^{m}\}\right) \bmod Y^{k+1}, 
\]
where $\alpha$ equals $\partial_Z(P)(0,a)$. 
\end{theorem}

\subsection{The Polishchuk--Spielman lemma}

\begin{lemma}[\cite{PolishchukS1994}]\label{lem:PS}
Let $\F$ be any field and $A(X,Y), B(X,Y) \in \F[X,Y]$ be polynomials with $X,Y$-degrees at most $D$. Let there be $n$-values $x \in \F$ such that $A(x,Y)$ divides $B(x,Y)$ in the ring $\F[Y]$ and the quotient $C_x(Y)$ has degree at most $k$, and let there be $n$-values $y \in \F$ such that $A(X,y)$ divides $B(X,y)$ in the ring $\F[X]$ and the quotient $R_y(X)$ has degree at most $k$. 

There exists constants $c, c' > 0$ with $c > c'$ such that if $n > cD$, then there is a bivariate $Q(X,Y)$ of degree at most $k$ in each variable such that 
\[
B(X,Y) = Q(X,Y)\cdot A(X,Y) .
\]
Furthermore, for all but $c'D$ choices of the substitutions $x, y$, $Q(x,Y) = C_x(Y)$ and $Q(X,y) = R_y(X)$. 
\end{lemma}

\subsection{Pruning to subgraph of high min degree}

\begin{lemma}\label{lem:pruning-prelims}
Let $G = (U, V, E)$ be a bipartite graph, such that the maximum degree of a vertex $u \in U$ is at most $\ell$, the maximum degree of a vertex $v \in V$ is at most $r$. Let $\epsilon, \delta$ be parameters such that the average degree of a vertex in $U$ is $\epsilon \ell$ and the average degree of a vertex in $V$ is $\delta r$. 

Then, there exists a subgraph $G' = (U', V', E')$ of the graph $G$ such that 
\begin{itemize}
    \item $|U'| \geq \frac{4\epsilon}{5}|U|$, $|V'| \geq \frac{4\delta}{5}|V|$.
    \item The degree of every vertex in $U'$ in the graph $G'$ is at least $\epsilon \ell/10$ and the degree of every vertex in $R'$ in the graph $G'$ is at least $\delta r/10$. 
\end{itemize}
\end{lemma}
\begin{proof}
To prove the lemma, we iteratively start deleting low-degree vertices from the graph $G$ as long as there is such a vertex. We argue that this procedure stops at some point, and when it stops, the remaining graph has the desired properties. 
\bigskip

\textbf{\underline{Algorithm Prune}}
\begin{itemize}
\item Set flag $\phi$ to $1$.
\item If $\deg(u) \geq \frac{\epsilon \ell}{10}$ for every $u \in U$, $\deg(v) \geq \frac{\delta r}{10}$ for every $v \in V $, 
            set $\phi$ to $0$ and stop. 
\item While $\phi$ is $1$: 
    \begin{enumerate}
        \item For $u \in U$:
            \begin{itemize}
                \item if $\deg(u) < \frac{\epsilon \ell}{10}$, delete $u$ and update $U \leftarrow U \setminus \{u\}$. 
                \end{itemize}

            \item For $v \in V$:
            \begin{itemize}
                \item if $\deg(v) < \frac{\delta r}{10}$, delete $v$ and update $V \leftarrow V \setminus \{v\}$. 
                \end{itemize}

                \item If $\deg(u) \geq \frac{\epsilon \ell}{10}$ for every $u \in U$, $\deg(v) \geq \frac{\delta r}{10}$ for every $v \in V $, 
            set $\phi$ to $0$ and stop.

    \end{enumerate}
\end{itemize}
\bigskip

Here, by the degree of a vertex, we mean the degree in the subgraph of $G$ alive at that moment, after any deletions that might have occurred in the algorithm. 

If $\phi$ equals one at the beginning of the \emph{While} loop, then at least one vertex is deleted from the graph in that run of the loop, and hence this process will terminate in finitely many steps. Note that at any step of the algorithm, a vertex in $U$ is deleted if its degree in the current graph is less than $\frac{\epsilon \ell}{10}$, and hence its deletion results in the deletion of at most $\frac{\epsilon \ell}{10}$ edges from the graph. Similarly, the deletion of any right vertex in the second for loop results in the deletion of at most $\frac{\delta r}{10}$ edges from the graph. 

Thus, the maximum number of edges that could have been deleted during the run of the algorithm is at most $\left( |U| \cdot \frac{\epsilon \ell}{10} + |V| \cdot \frac{\delta r}{10} \right)$, which is at most $\left( \frac{|E|}{5}\right)$, since $\epsilon \ell |U| = \delta r |V| = |E|$. So at least $\left( \frac{4 |E|}{5}\right)$ edges must have survived the pruning process. 

Hence, the graph $G'$ obtained at the end of the process is non-empty. Clearly, the minimum degree condition on the vertices will be satisfied at this point. The lower bound on the number of surviving left and right vertices now follows from the number of surviving edges and the bounds on the maximum degrees. 

\end{proof}

\section{Consistent roots on parallel lines to global rational roots}\label{sec:lifting-lemma}
In this section, we state and prove the following technical lemma that is crucial for the proof of the generalization of the Polishchuk--Spielman lemma in \autoref{thm:PS-gen-intro} as well as  to the proof of \autoref{thm:main technical}.

\consistentparallellineslift*

For the proof of the lemma, we first do an affine change of basis and reduce to the setting where the lines in $\calL$ are parallel to the $Y$-axis and the line $\hat{L}$ is the $X$-axis. This helps simplify the notation in the proof while conveying most of the main technical ideas. 

\subsection{The case of axis parallel lines}
\begin{lemma}\label{lem:lifting-parallel-lines}
Let $Q(X,Y,Z)$ be a non-zero trivariate polynomial of $(1,1,k)$-degree $D$. Let $S \subseteq \F$ be a set of field elements such that for every $a \in S$, there exists a polynomial $g_a(Y)$ of degree at most $k$ that satisfies 
 \[
 Q(a,Y,g_a(Y)) \equiv 0 . 
 \]
 Furthermore, let $h(X)$ be a degree $k$ polynomial such that $Q(X,0,h(X))$ is identically zero, $\partial_Z(Q(X,0,h(X)))$ is non-zero, and for every $a \in S$, $h(a) = g_a(0)$. 

 If $\left(|S| > 2D + d_Z(Q)D(2k+1)\right)$, then there exists $U(X,Y) \in \F(X)[Y]$ such that $\deg_Y(U) \leq k$, $\deg_X(U) \leq Dk$ and \[
 Q(X,Y,U(X,Y)) \equiv 0. 
 \] 
 Moreover, for all but at most $D$ choices of $a \in S$, $U(a,Y) = g_a(Y)$. 
 \end{lemma}
\begin{proof}
    From the bound on the weighted degree of $Q$, we have that the degree of $\partial_Z(Q(X,0,h(X)))$  is also at most $D$. Moreover, since $\partial_Z(Q(X,0,h(X)))$ is a non-zero polynomial of degree at most $D$, it can have no more than $D$ zeroes in the set $S \subseteq \F$. Let $\hat{S}$ be the subset of $S$ defined as 
    \[
    \hat{S} = \left\{a : a \in S \text{ and } \partial_Z(Q(a,0,h(a))) \neq 0 \right\}.
    \]
Thus, we have that 
\[
|\hat{S}| \geq |S| - D .
\]
    We now view $Q$ as a bivariate in $Y, Z$, with coefficients over the field $\F(X)$. From the hypothesis of the lemma, we have that 
    \begin{align*}
        Q(X,Y,h(X)) &= 0 \bmod Y, \quad \text{and} \\
        \partial_Z(Q(X,Y,h(X))) &\neq 0 \bmod Y. 
    \end{align*}
    Thus, from \cref{thm:NI}, we have that there is a unique power series $\Phi(X,Y)$ in $Y$, with coefficients from the field $\F(X)$ such that 
    \begin{align*}
        &Q(X,Y,\Phi(X,Y)) \equiv 0 \quad \text{and}\\
        &\Phi(X,Y) \equiv h(X) \bmod Y. 
    \end{align*}
    Let $U(X,Y)$ be defined as the truncation of $\Phi(X,Y)$ modulo $Y^{k+1}$. Thus, $U(X,Y)$ is a univariate in $Y$ of degree $k$ with coefficients in the field $\F(X)$ of rational functions in $X$, and is the unique such element that satisfies 
    \begin{align*}
    &Q(X,Y,U(X,Y)) \equiv 0 \bmod Y^{k+1} \quad \text{and}\\
    &U(X,Y) \equiv h(X) \bmod Y. 
    \end{align*}
To complete the proof of the lemma, we need the following two claims. We first complete the proof of the lemma using the claims and then discuss the proofs of the claims. 

    \begin{claim}\label{clm:U-restricts-to-true-roots}
    For every $a \in \hat{S}$,  the denominator of $U(a,Y)$ is non-zero and $U(a,Y) = g_a(Y)$. In particular, 
    \[
    Q(a,Y,U(a,Y)) \equiv 0. 
    \]
    \end{claim}

    \begin{claim}\label{clm:X-deg-U}
    
$    \deg_X(U) \leq D(2k + 1).$
    \end{claim}
To prove the theorem, we consider the formal rational function $R(X,Y) := Q(X,Y,U(X,Y))$. Clearly, $R$ is a polynomial in $Y$ and a rational function in $X$. Moreover, $R$ can be written as 
\[
R = \sum_{i = 0}^{d_Z(Q)} Q_i(X,Y) U(X,Y)^i , 
\]
where each $Q_i$ is a polynomial of total degree at most $D$ (in fact, at most $(D - d_zk))$. Thus, $\deg_X(R)$ is at most $D' := (D + d_z\cdot \deg_X(U))$. In particular, let us focus on the numerator of the $R$, whose degree is at most $D'$. From \cref{clm:U-restricts-to-true-roots}, we have that for every $a \in \hat{S}$, $R(a,Y)$ is well defined (its denominator does not vanish), and its numerator evaluates to zero. Thus, if 
\[
|\hat{S}| > D', 
\]
we get that $R(X,Y)$ is identically zero, and $U(X,Y)$ is a $Z$-root of $Q$. Moreover, from \cref{clm:U-restricts-to-true-roots}, we get that for every $a \in \hat{S}$, $U(a,Y)$ equals $g_a(Y)$.  

But the inequality needed on the size of $\hat{S}$ immediately follows from $|\hat{S}| \geq (|S| - D)$, the upper bound on $\deg_X(U)$ in \cref{clm:X-deg-U}. 
\end{proof}

\paragraph*{Proof of technical claims}
\begin{proof}[Proof of \cref{clm:U-restricts-to-true-roots}]
From \cref{thm:NI-explicit}, we have that the denominator of $U(X,Y) \in \F(X)[Y]$ is a power of $\partial_Z(Q(X,0,h(X)))$. Thus, from the definition of $\hat{S}$, we get that for every $a \in \hat{S}$, this denominator is non-zero. 

Since $Q(X,Y,U(X,Y))$ is zero modulo $Y^{k+1}$, we get that for every $a \in \hat{S}$, $U(a,Y)$ is well defined, and \[Q(a,Y,U(a,Y)) \equiv 0 \bmod Y^{k+1}.  \]
Moreover, the degree of $U(a,Y)$ is at most $(k+1)$ and the constant term $U(a,0)$ of $U(a,Y)$ equals $h(a)$. From the hypothesis of the lemma, this equals $g_a(0)$. 

Thus, $U(a,Y)$ and $g_a(Y)$ are both degree $k$ polynomials in $Y$ that are the \emph{lifts} of the root $h(a)$ of $Q(a,0,Z)$, which additionally satisfies $\partial_Z(Q)(a,0,h(a)) \neq 0$. By the uniqueness part of \cref{thm:NI}, we get that $U(a,Y)$ must equal $g_a(Y)$, and thus, $Q(a,Y,U(a,Y))$ must be identically zero. 
\end{proof}

\begin{proof}[Proof of \cref{clm:X-deg-U}]
    The claim immediately follows from the expression for approximate $Z$-roots of the bivariate $Q(X,Y,Z) \in \F(X)[Y,Z]$ in \cref{thm:NI-explicit}, when invoked with  initial conditions $Q(X,0,h(X))$ is zero  and $\partial_Z(Q(X,0,h(X)))$ is non-zero. 
    
\end{proof}

\subsection{From axis parallel to the general case }
We now complete the proof of \autoref{lem:lifting-parallel-lines-general}. 
\begin{proof}[Proof of \autoref{lem:lifting-parallel-lines-general}]
For the ease of notation, we assume that neither the line $L$ nor the lines in $\calL$ are parallel to the $Y$-axis. The argument below can be easily extended to this case by a different parameterization of the lines. 

Let $L_0$ be the line $L_0 := \{(x,mx) : x \in \F\}$ with slope $m \in \F$ that is parallel to the lines in $\calL$. Similarly, let $\hat{L}$ be the line $\hat{L} := \{(x,\hat{m}x+\hat{u}) : x \in \F\}$ for some $\hat{m},\hat{u}$ in $\F$. We also know that $\hat{m} \neq m$. Let $V,W$ be new formal variables and let us consider the system of linear constraints
\begin{align}\label{eqn:basis-change}
    Y - mX &= V \\ 
    Y - \hat{m}X &= W+\hat{u} \, .
\end{align}
Since $m \neq \hat{m}$, this system has a unique solution and gives us a way to write $X$ and $Y$ as affine linear combinations of $V,W$, namely $X=\frac{(V-W-\hat{u})}{(\hat{m}-m)}$ and $Y=\frac{(\hat{m}V-m(W+\hat{u}))}{(\hat{m}-m)}$. 

Let $P(V,W, Z) \in \F[V,W,Z]$ be a trivariate polynomial defined as
\[
P(V,W,Z) := Q\left(\frac{(V-W-\hat{u})}{(\hat{m}-m)}, \frac{(\hat{m}V-m(W+\hat{u}))}{(\hat{m}-m)}, Z \right) \, .
\]
Clearly, $P(Y-mX, Y-\hat{m}X-\hat{u}, Z)$ equals $Q(X,Y,Z)$. 

Now, for any line $L$ that is parallel to $L_0$, parameterized as $L := \{(x,mx+{u}) : x \in \F\}$ and polynomial $g_L(X)$, 
\[
Q(X,mX+u,g_L(X)) = P(u, X(m-\hat{m})+(u-\hat{u}),g_L(X)).
\] 
Furthermore, by setting ${W} := (m-\hat{m})X+(u-\hat{u})$, we get a polynomial $\tilde{g}_L$ with degree at most the degree of $g_L$ such that if $Q(X,mX+u,g_L(X))$ is identically zero, then $P(u,{W},\tilde{g}_L({W}))$ is also identically zero. 

Similarly, the degree-$k$ $Z$-root $\hat{h}$ of $Q((X,Y)_{\hat{L}},Z)$ implies that $P({V},0,\tilde{h}({V}))$ is identically zero for a polynomial $\tilde{h}$ of degree at most $k$. 

Thus, the polynomial $P$ satisfies the hypothesis of \autoref{lem:lifting-parallel-lines}. From the conclusion of the lemma, we get a rational function $U(V,W) \in \F(V,W)$ with numerator and denominators of degree at most $Dk$ such that $U$ is a $Z$-root of $P$. Now, writing $V, W$ as linear combinations of $X,Y$ as in equation \autoref{eqn:basis-change}, we get a rational function $\tilde{U}(X,Y)$ of degree at most $Dk$ that is a $Z$-root of $Q$. Note that this transformation between $X,Y$ and $V,W$ is invertible, so the non-zero denominator of $U$ cannot vanish under this transformation. 

The moreover part also follows immediately from the basis change and the guarantees on the restrictions of $U$ in \autoref{lem:lifting-parallel-lines}. 
\end{proof}

\section{Generalizations of the Polishchuk--Spielman lemma}\label{sec:PS-gen}
We now state and prove the generalization of \autoref{lem:PS} that drives the analysis of our low-degree test. 
\begin{theorem}\label{thm:gen-PS-parallel-lines}
    There exists a constant $c > 1$ such that for all $D,k,q, p, r \in \N$ where $q$ is a prime power, $\F$ is the finite field of size $q$ and $Q(X,Y,Z) \in \F[X,Y,Z]$ is a polynomial with $(1,1,k)$-degree $D$ 
     the following is true. 
    
    Let $\calL_1, \dots, \calL_r$ be subsets of lines in $\F^2$ such that for every $i \in [r]$, $\calL_i$ is the set of all $q$ parallel lines in a certain direction, the lines in $\calL_i$ and $\calL_j$ are intersecting for $i\neq j$. Furthermore, for every $i \in [r]$ and line $L \in \calL_i$, there is a polynomial $g_L$ of degree at most $k$ that is a $Z$-root of $Q((X,Y)_L,Z)$.
    
    If $r > d_Z(Q)$, and $q > c\poly(\deg_Z(Q))Dk$, then there is a bivariate polynomial $P(X,Y)$ of total degree at most $2k$ so that $Q(X,Y,P(X,Y))$ is identically zero and there are
    distinct indices $i_0, j_0 \in [r]$ such that the restriction of $P$ on any line $L$ equals $g_L$ for all but $O(Dk)$ choices of lines $L$ in $\calL_{i_0} \cup \calL_{j_0}$. 

\end{theorem}

In the rest of this section, we prove this theorem.

\subsection*{Proof of \autoref{thm:gen-PS-parallel-lines}}
The rest of this section is devoted to the proof of \autoref{thm:gen-PS-parallel-lines}. The proof is in divided into various stages. We first start by preprocessing $Q$ in a way that ensures additional structure on the problem instance without loss of generality. We then zero-in on two of the $r$-directions that are \emph{good} in some formal sense. Using the lines in these two directions, we conclude that $Q$ must have a $Z$-root that is a rational function in $X,Y$. This is the step that involves the use of the generalization of the Polishchuk--Spielman lemma in \autoref{lem:lifting-parallel-lines-general}. Finally, we conclude that the root obtained in the previous step is not just a rational function; it is a polynomial of the appropriate degree, and concludes the proof of \autoref{thm:gen-PS-parallel-lines}. 

\subsection{Preprocessing}
We start with the preprocessing step here. 
\subsubsection*{The square-free part}
Let $Q_1(X,Y,Z)$ be the square-free part of $Q$ (\autoref{def:square-free-part}). Then, clearly, $Q_1$ also satisfies the hypothesis of the theorem. Furthermore, if the conclusion is true for $Q_1$, then it is also true for $Q$. So, we work with $Q_1$ henceforth. 

\subsubsection*{$Z$-degree at least one}\label{sec:Z-deg-1}
The hypothesis of the theorem may be true on some of the lines $L$ because $Q_1((X,Y)_L,Z)$ is identically zero. From \autoref{lem:linear-factors}, we know that the total number of such lines is at most the total degree of $Q_1$, which is at most $D$. In fact, a stronger conclusion is true here: for all but at most $D$ lines $L$, the $Z$-degree of $Q_1((X,Y)_L,Z)$ equals the $Z$-degree of $Q_1$. We discard the $O(D)$ \emph{bad lines} (note that the number of lines $q$ in each direction is large enough compared to $D$ for us to be able to tolerate this loss), and for the rest of the proof assume that $Q_1((X,Y)_L,Z)$ is non-zero and equals the $Z$-degree of $Q_1(X,Y,Z)$ for all lines $L$ under consideration. In particular, this ensures that the $Z$-degree of $Q_1$ is at least one for the hypothesis to hold.

Furthermore, we will assume without loss of generality that for at least one of the line polynomials $g_L$, the degree equals $k$, else we can replace $k$ everywhere by $(k-1)$. This also implies that $k\leq D$, since otherwise a degree $k$ polynomial $(Z-g_L)$ cannot be a factor of the polynomial $Q_1((X,Y)_L,Z)$ of total degree at most $D$. 

\subsubsection*{Ensuring a non-zero $Z$-derivative}
This step of pre-processing can be skipped if the underlying field $\F$ has characteristic either zero or larger than the $Z$-degree of $Q_1$, since over such fields $\deg_Z(Q_1) \geq 1$ implies that $\partial_Z(Q_1)$ is a non-zero polynomial. Some of the details are simpler in this setting, and we encourage the reader to read the proof with this setting in mind for the first time. 

Over fields $\F$ of small positive characteristic, it may be the case that $\deg_Z(Q_1) \geq 1$ and yet $\partial_Z(Q_1)$ is identically zero. From \autoref{lem:hasse-props}, we know that this happens if and only if $Q_1$ is a polynomial in $Z^{p}$ for some $s \geq 1$. Let $s \in \N$ be  such that $Q_1$ is a polynomial in $Z^{p^s}$ and not in $Z^{p^{s+1}}$. From \autoref{lem:hasse-props}, we get that there exists a polynomial $\hat{Q}(X,Y,Z)$ and a natural number $s \geq 1$ such that $\partial_Z(\hat{Q})$ is non-zero, and $Q_1(X,Y,Z) = \hat{Q}(X,Y,Z^{p^s})$. For the next few steps of the proof, we work with $\hat{Q}$. We eventually return to $Q_1$ towards the end. 

To proceed further, we first discuss the properties of $\hat{Q}$ that follow from the properties of $Q_1$. Since the $(1,1,k)$-degree of $Q_1$ is at most $D$, we get that the $(1,1,p^sk)$-degree of $\hat{Q}$ is at most $D$. Furthermore, for any line $L \in \F^2$, if $g_L$ is a degree $k$ polynomial such that $Q_1((X,Y)_L,g_L)$ is identically zero, then we get 
\[
Q_1((X,Y)_L,g_L) = \hat{Q}((X,Y)_L,g_L^{p^s}) \equiv 0 \, .
\]
 Thus, ${g_L}^{p^s}$ is a $Z$-root of $\hat{Q}((X,Y)_L,Z)$ for any line $L$ where $g_L$ is the $Z$-root of $Q_1((X,Y)_L,Z)$. So, the property of having $Z$-roots on lines is preserved, although the degree of each of the $Z$-roots on lines is now bounded above by $p^sk$. 
 We also record the following simple observation that follows from the definitions so far. 

 \begin{observation}\label{obs:props-of-Q-hat}
 $\hat{Q}$ is square-free,  $\partial_Z(\hat{Q})$ is non-zero. If line $L$ is such that 
 \[
 \deg_Z(Q_1(X,Y,Z)) = \deg_Z(Q_1((X,Y)_L,Z)), 
 \]
 then,  
 \[
 \deg_Z(\hat{Q}(X,Y,Z)) = \deg_Z(\hat{Q}((X,Y)_L,Z)).
 \]
 \end{observation}
 
Going forward, we work with $\hat{Q}$, whose $(1,1,kp^s)$-degree is at most $D$, and has $Z$-roots of degree at most $\hat{k} = kp^s$ on many lines. Furthermore, each of these roots is the $p^{s}$-th power of a polynomial of degree at most $k$. 

\subsection{Finding two good directions}
In this phase, we use the fact that the number of directions $r$ is strictly greater than the $d_Z := \deg_Z({Q}) \geq \deg_Z(\hat{Q})$ (and hence $\deg_Z(\hat{Q})$) to conclude that there is a pair of directions $i,j \in [r]$ such that the line polynomials for lines in $\calL_i$ and $\calL_j$ have non-trivial consistency. This is formalized in the following lemma. 

\begin{lemma}\label{lem:two-good-directions}
There exist $i_0, j_0 \in [r]$ such that $i_0 \neq j_0$ and there is a set $\mathcal{P}$ of points in $\F^2$ of size at least $\Omega(q^2/d_Z^2)$ such that for every point $\va \in \mathcal{P}$ and lines $L \in \calL_{i_0}$ and $L' \in \calL_{j_0}$ incident on $\va$, the corresponding \emph{line polynomials} $g_{L}$ and $g_{L'}$ agree on $\va$. 
\end{lemma}
\begin{proof}
Let $\va \in \F^2$ be any point and let $L_1, L_2, \dots, L_{r}$ be the lines incident on $a$ such that $L_i \in \calL_i$. Let $g_{L_i}$ be the corresponding line polynomials such that $\hat{Q}((X,Y)_{L_i},{g}_{L_i})$ is identically zero. Since each of the lines $L_i$ is incident to $\va$, we get that $\hat{Q}(\va, g_{L_i}(\va))$ equals zero, or equivalently, $g_{L_i}(\va)$ is a zero of the univariate $\hat{Q}(\va, Z)$. But $\hat{Q}(\va, Z)$ has degree at most $d_Z$, and $r > d_Z $ so there is a subset of size at least  $\lceil r/d_Z \rceil \geq 2$ of these lines such that the corresponding line polynomials agree on $\va$. 

Thus, by Pigeonhole principle, there is a pair of directions ${i_0,j_0} \in [r]$ such that the line polynomials in these directions agree on at least $\frac{\binom{\lceil r/d_Z \rceil}{2}}{\binom{r}{2}}$ fraction of all points $\va$. For $r > \deg_Z(Q) \geq 1$, this fraction is at least $\frac{1}{2d_Z^2}$. 

Finally, we note that we can only do the above arguments on points in $\F^2$ that have a line from each of the $r$ sets $\calL_1, \ldots, \calL_r$ incident on it. In particular, we need to exclude all points in $\F^2$ that may be incident to any of the $O(D)$ bad lines excluded in \autoref{sec:Z-deg-1}. Thus, the size of the set $\mathcal{P}$ of points where lines in $\calL_{i_0}, \calL_{j_0}$ are consistent with each other is at least $\frac{q(q-O(Dd_Z^2))}{2d_Z^2}$, which, for $D < q/(cd_Z^2)$ (for a sufficiently large constant $c$) is $\Omega(q^2/d_Z^2)$. 
\end{proof}

\subsection{Rational roots from $\calL_{i_0}$ and $\calL_{j_0}$}
For brevity of notation, we assume that $i_0 = 1$ and $j_0 = 2$. From \autoref{lem:two-good-directions}, we know that line polynomials on lines in $\calL_1$ and $\calL_2$ are consistent on at least $\alpha q^2/d_Z^2$ points in $\F^2$ for some absolute constant $\alpha > 0$. Let $G$ be a bipartite graph with vertices $\calL_1, \calL_2$ with there being an edge between lines $L \in \calL_1, L' \in \calL_2$ if the corresponding line polynomials $g_L$ and $g_{L'}$ are consistent, i.e. they agree on the point $L \cap L'$. Clearly, both sets of vertices in the bipartition have size at most $q$, the maximum degree of any vertex is $q$ and there are at least $\alpha q^2/d_Z^2$ edges in the graph. 

From \autoref{lem:pruning-prelims}, we get that there is a subgraph $G'$ of $G$ with vertex sets $\calH_1 \subseteq \calL_1$ and $\calH_2 \subseteq \calL_2$ of size at least $\frac{ 4\alpha}{5} q/d_Z^2$ each and minimum degree at least $\frac{ \alpha}{10} q/d_Z^2$. 

We focus on lines in this subgraph $G'$ now, and refer to the lines in $\calH_1, \calH_2$ as \emph{heavy} lines.   

\subsubsection*{Setting up an anchor}

We now prove the following crucial technical claim. 
\begin{claim}\label{clm:good-anchor}
For all but $O(d_ZD)$ lines $L$ in  $\calH_1 \cup \calH_2$, $\partial_Z(\hat{Q}((X,Y)_{L}, g_{L}))$ is non-zero. 
\end{claim}
\begin{proof}
Recall that $\hat{Q}$ has the property that $\partial_Z(\hat{Q})$ is a non-zero polynomial. Furthermore, $\hat{Q}$ is square-free. We know from \autoref{lem:linear-factors} that there are at most $\deg(\partial_Z(\hat{Q})) \leq D$ lines $L$ in total such that  $\deg(\partial_Z(\hat{Q})((X,Y)_L,Z))$ is strictly smaller than $\deg(\partial_Z(\hat{Q})(X,Y,Z)$ or $\partial_Z(\hat{Q})((X,Y)_L,Z)$ is identically zero. Let $B_1$ be the subset of all such $O(D)$ lines. 

Furthermore, $D < \frac{q}{ck\poly(d_Z)}$ for some $c > 1$, so, there exist lines in $\calH_1$ and in $\calH_2$ that are not in $B_1$. 

If the $Z$-degree of $\partial_Z(\hat{Q})$ is zero, we get that for any such $L$ not in $B_1$, $\partial_Z(\hat{Q}((X,Y)_{L_i}, g_{L_i}))$ is non-zero and we are done.

Else, the $Z$-degree of $\partial_Z(\hat{Q})$ is at least one. Since $\hat{Q}$ is square-free, we get from \autoref{thm:disc-and-sqr-free-multivariates} that $\disc_Z(\hat{Q})$ is a non-zero bivariate polynomial, and has degree at most $2d_ZD$. Thus, from \autoref{lem:linear-factors}, there are at most $2d_ZD$ lines $L$ such that $\disc_Z(\hat{Q})$ vanishes identically when restricted to these lines. Let $B_2$ be the set of such lines. Clearly, $|B_2| \leq 2d_ZD$. 

From \autoref{obs:resultant-restrictions-commute}, we get that for any line $L$ that is not in $B_1 \cup B_2$, the polynomials $\hat{Q}((X,Y)_L, Z)$ and $\partial_Z(\hat{Q}((X,Y)_{L}, Z))$ do not share a common $Z$-root, and if the polynomial $g_L$ is a root of $\hat{Q}((X,Y)_L, Z)$, then it cannot be a root of $\partial_Z(\hat{Q}((X,Y)_{L}, Z))$. 

The claim now follows from the upper bound on the sizes of $B_1$ and $B_2$. 
\end{proof}

\subsubsection*{Existence of a rational root}

Let $L_1 \in \calH_1$ be a line not excluded in \autoref{clm:good-anchor}. Since $L_1$ is heavy, the line polynomial on $L_1$ is consistent with at least $\Omega(q/d_Z^2)$ of the line polynomials for lines in $\calH_2$. Let $S_1$ be the subset of $\calH_2$ of these lines. We also know from \autoref{clm:good-anchor} that $\partial_Z(\hat{Q}((X,Y)_{L_1}, g_{L_1}))$ is non-zero and $\hat{Q}((X,Y)_{L_1}, g_{L_1})$ is identically zero. Recall that the $(1,1,\hat{k})$-degree of $\hat{Q}$ is at most $D$ and each of the line polynomials has degree at most $\hat{k}$, where $\hat{k} \leq kp^s$. So, a stronger bound on $\deg_Z(\hat{Q})$ is that $\deg_Z(\hat{Q}) \leq \deg_Z(Q)/p^s \leq D/\hat{k} = D/(kp^s)$. 

From the setting of parameters, we have that $|S_1| \geq \Omega(\poly(\deg_Z(Q)) D{k})$, for $q > c\poly(d_Z)Dk$ for a sufficiently large constant $c$. This implies that 
\[
|S_1| \geq \Omega(\deg_Z(\hat{Q})D\hat{k}). 
\]
Thus, from \autoref{lem:lifting-parallel-lines-general}, we get that there is a rational function $U_1(X,Y)$ of degree at most $D\hat{k}$ such that 
\[
\hat{Q}(X,Y,U_1(X,Y)) \equiv 0 , 
\]
and for all but $D$ choices of lines $L \in S_1$, $U_1((X,Y)_{L}) = g_{L}$. 

 Let $S_1'$ be the set of lines $L$ in $S_1$ such that $U_1((X,Y)_{L}) = g_{L}$. We have that $|S_1'| \geq |S_1| \geq \Omega(q/\deg_Z(\hat{Q})^2)$ since $D < \frac{q}{ck\poly(\deg_Z(Q))} \leq \frac{q}{c\hat{k}\poly(\deg_Z(\hat{Q}))}$. Let us consider the induced subgraph of $G'$ on vertices $S_1' \cup \calH_1$. 
 
 The degree of vertex in $S_1'$ in this graph is at least $\frac{\alpha q}{10\deg_Z(\hat{Q})^2}$ and hence the number of edges is at least $\frac{\beta q^2}{\poly(\deg_Z(\hat{Q}))}$ for some constant $\beta > 0$. Once again, by an application of \autoref{lem:pruning-prelims}, we get that this graph has an induced subgraph $G'' = (\calH_1', S_1'', E'')$ where the size of the left and right side of the bipartition as well as the minimum degree are at least $\Omega\left(\frac{q}{\poly(\deg_Z(\hat{Q}))}\right)$. 

 We now repeat the argument  applied to the graph $G'$ over the last two sections to this new graph $G''$: we pick a line  $L_2$ in $S_1''$ as the anchor (using \autoref{clm:good-anchor} again), and pick its neighbors in $G''$ as in the set $S_2$ and invoke \autoref{lem:lifting-parallel-lines-general} to obtain a rational function $U_2(X,Y)$ of degree at most $Dk$ whose restriction on all but $D$ lines $L$ in $\calH_1'$ equals the corresponding line polynomial $g_L$. Let $\calH_1''$ be the subset of $\calH_1'$ whose line polynomials are consistent with $U_2$. Note that the number of edges in $G''$ with both end points in $S_1'' \cup \calH_1''$ is at least $\Omega(q^2/\poly(\deg_Z(\hat{Q})))$, where once again we use the fact that $Dk$ (and hence $D$) is much smaller than $q$. 

Recall that $S_1'' \subseteq S_1'$ and hence, for every line $L$ in $S_1''$, $U_1((X,Y)_L) = g_L$. This means that for every $L' \in \calH_1''$ and $L \in S_1''$, if $L'$ and $L$ are adjacent in $G''$, then $U_1$ and $U_2$ agree on the point $L \cap L'$. Thus, $U_1$ and $U_2$ are rational functions of degree $D\hat{k}$ and agree on at least $\Omega(q^2/\poly(\deg_Z(\hat{Q})))$ many points in $\F^2$. Since $D\hat{k} < q/\poly(\deg_Z(\hat{Q}))$, this means, from \autoref{lem:SZ}, that $U_1$ and $U_2$ must be the same rational function. 

\subsection{From a rational function to a polynomial}
Let $U$ denote the rational function in simplest form that equals $U_1, U_2$ constructed in the previous subsection. From the construction above, we know that its degree is at most $D\hat{k}$, and it equals a degree $\hat{k}$ polynomial when restricted to $\Omega(q/\poly(\deg_Z(\hat{Q})))$ many lines in two distinct directions. Since $D\hat{k}$ is much smaller than the number of lines in each direction, we apply the classical Polishchuk--Spielman lemma (\autoref{lem:PS}) on the numerator and the denominator of $U$ to conclude that there is a bivariate polynomial $P(X,Y)$ of degree at most $2\hat{k}$ that equals $U$ on all but $O(D)$ of the $\Omega(q/\poly(\deg_Z(\hat{Q})))$ lines in consideration in the two directions.

\subsection{A $Z$-root of $Q$}

If the characteristic $p$ of the underlying field $\F$ exceeds $\deg_Z(Q)$, then the polynomial $\hat{Q}$ is equal to $Q_1$, and hence $P(X,Y)$ is a $Z$-root of $Q_1$, and hence of $Q$, and this completes the proof of the theorem.

Over fields of (small, positive) characteristic $p$, we have that $Q_1(X,Y,Z) = \hat{Q}(X,Y,Z^{p^s})$ for some positive integer $s$. From the discussion so far, we know that $P(X,Y)$ is a $Z$-root of $\hat{Q}$ of in other words, $(Z-P(X,Y))$ divides $\hat{Q}(X,Y,Z)$. Substituting $Z$ by $Z^{p^s}$, we get that $(Z^{p^s} - P(X,Y))$ divides $\hat{Q}(X,Y,Z^{p^s})$, and hence $Q_1(X,Y,Z)$ and therefore the original polynomial $Q(X,Y,Z)$. To complete the proof of the theorem, it suffices to show that $P(X,Y)$ equals $\hat{P}(X,Y)^{p^s}$ for some polynomial $\hat{P} \in \F[X,Y]$. 

To see this, we recall that $P$ has the property that there exist two distinct directions in $\F^2$ such that the restriction of $P$ on at least $\Omega(q/\poly(\deg_Z(\hat{Q})))$ lines in these directions equals the line polynomial $g_L$. But from the relationship between $\hat{Q}$ and $Q_1$, we know that $g_L = \tilde{g}_L^{p^s}$, where $\tilde{g}_L$ is the $Z$-root of $Q_1((X,Y)_{L},Z)$. 

Let $s' \leq s$ be the largest integer such that there is a polynomial $\hat{P}$ satisfying $P = \hat{P}^{p^{s'}}$, and $\hat{P}$ is not the $p$th power of any polynomial in $\F[X,Y]$. Now, if $s' < s$, then $\hat{P}$ must be the $p$th power of every line polynomial  $g_L$ for which $g_L^{p^s} = P_L$. But, we have this property for more than $\deg(P)$ lines in two different directions, and hence by \autoref{lem:pth-power-on-lines}, we get that $\hat{P}$ is also a $p$th power, which is a contradiction. Thus, $s'$ must equal $s$ and $\hat{P}$ and $P$ satisfy $P = \hat{P}^{p^s}$. 

So, we get that $(Z^{p^s} - \hat{P}^{p^s}) = (Z-\hat{P})^{p^s}$ divides $Q_1$ and hence $Q$. This completes the proof of \autoref{thm:gen-PS-parallel-lines}.

\section{The low-degree test}\label{sec:LDT-analysis}
In this section, we analyze the derandomized version  of the line-point low-degree test in the high-error regime. We start by recalling the set up. 

\paragraph{Setup: }
Let $\calU = \{u_1, u_2, \ldots, u_r\} \subseteq \F^2$ be \emph{directions} of interest to us. $u_i's$ represent distinct directions and hence, $u_i$ and $u_j$ are not scalar multiples of each other for any $i \neq j$. Let $\calL_1, \dots, \calL_r$ be subsets of lines in $\F^2$ such that for every $i$, $\calL_i$ is the set of \emph{all} lines in the direction $u_i$ in $\F^2$. Clearly, $|\calL_i| = |\F| = q$. Clearly, for every point $\va \in \F^2$, and for every $i \in [t]$, there is a unique line $L_{i,v}$\footnote{note that the same line can be indexed in many different ways} in $\calL_i$ that is incident to $\va$. Let $\calL$ denote the union of $\calL_i$. 

We now restate the theorem we wish to prove. 
We recall the main theorem. 
\LDTanalysismain*

Before we proceed it will be convenient to understand the hypothesis of the above theorem, namely 
\[ 
\E_{\ell \in \calL}[\agr(f|_\ell, \calP_{k,1})] \geq \eps\,.
\]
It will be convenient to interpret this statement in terms of a subgraph of the points-vs-line incidence graph. The points-vs-line incidence graph is the graph between the set $\mathcal{P}$ of all points in $\F^2$ and set $\calL$ of allowed lines where a point $\va \in \mathcal{P}$ is adjacent to a line $\ell in \calL$ if $\va \in \ell$. Let $E^*$ be the set of edges of this incidence graph. We  label the points $\mathcal{P}=\F^2$ with the function $f \colon \F^2 \to \F$, more precisely $\va \in \F^2$ is labeled with $f(\va)$. We label the lines with univariate polynomials of degree $k$ that have the best agreement with the restriction of the function $f$ on that line. More precisely, for each $\ell \in \calL$, we consider the restricted function $f|_\ell$ and find the univariate polynomial $P_\ell$ of degree at most $k$ that has the best agreement with $f|_\ell$ (if there exists more than one best-fit, we break ties arbitrarily). We now consider the subgraph $(\mathcal{P},\calL, E)$ of the points-vs-line incidence graph where $E \subseteq E^*$ is defined as follows:
\[ E = \{ (\va,\ell) \in \mathcal{P}\times \calL \mid \va \in \calL, P_\ell(\va)=f(\va)\}\,.
\]
In other words, $(\va,\ell)$ is an edge in $E$ iff the line polynomial $P_\ell$ agrees with the points table $f$ at the point $\va$. It immediately follows from the definition of this graph that 
$\E_{\ell \in \calL}[\agr(f|_\ell, \calP_{k,1})] \geq \eps$ iff  $|E| \geq \eps |E^*|$.

\subsection{Pruning to \emph{good} lines, \emph{good} points}
Consider the bipartite lines-vs-points incidence graph $G_f = (\mathcal{P},\calL,E)$ for the function $f$ defined as follows. The set $\mathcal{P}$ is the set of all points in $\F^2$ and the set $\calL$ is the union of all lines in $\cup_{i \in [r]} \calL_i$. For a point $\va \in \mathcal{P}$ and a line $L$ in $\calL$, $(v,L)$ is an edge in $G_f$ iff the line $L$ is incident to $\va$ and the line-point test for the function $f$ passes on the line-point pair $(L,\va)$. 

Clearly, the number of points is $q^2$, the number of lines is $qr$ and the maximum degree of any line in the graph $G_f$ is at most the field size, i.e. $q$ and the maximum degree of any point is $r$. Furthermore, if the test passes with probability $\epsilon$, then the graph $G_f$ has $\epsilon q^2r$ edges. Thus, the average degree of a line vertex in $G$ is $\epsilon q$ and the average degree of a point vertex is $\epsilon r$. 

Applying \autoref{lem:pruning-prelims} to the graph $G_f$ gives us the following corollary. 

\begin{corollary}\label{cor:pruning the incidence graph}
Let $f:\F^2 \to \F$ be a function such that $\E_{\ell \in \calL}[\agr(f|_\ell,\calP_{k,1})]$. Then, there exists a subgraph $G'_f = (\mathcal{P}', \calL', E')$ of the graph $G_f$ such that 
\begin{itemize}
    \item $|\mathcal{P}'| \geq \frac{4\epsilon}{5}q^2$, $|\calL'| \geq \frac{4\epsilon}{5}qr$.
    \item The degree of every vertex in $\mathcal{P}'$ is at least $\epsilon r/10$ and the degree of every vertex in $\calL'$ is at least $\epsilon q/10$. 
\end{itemize}

\end{corollary}

For the rest of this proof, we focus our attention on the lines and points in the graph $G'_f$. We refer to them as \emph{good} lines and \emph{good} points respectively. For the ease of notation, we also drop the subscript $f$ and simply refer to $G'_f$ as $G'$. 

\subsection{Interpolation}
From this section onwards, we will use $\hat{k}$ to denote the maximum of $k$ and $\log^2 q$. 
We think of $\hat{k}$ as a proxy for the degree parameter $k$.

In this section, we show that there is a non-zero trivariate polynomial $Q(X,Y,Z)$ of sufficiently low degree that in some sense succinctly \emph{captures} the fact that the low-degree test passes with probability at least $\epsilon$. This step is similar in spirit to the interpolation step in prior works, most notably in the low-degree test of Arora-Sudan \cite{AroraS2003}, and was subsequently used in the work of Harsha, Kumar, Saptharishi and Sudan \cite{HarshaKSS2024}. However, in our opinion, the details of the interpolation here are substantially simpler than those in these prior works. Furthermore, the quantitative guarantees are also somewhat stronger, and in particular the interpolated polynomial $Q$ \emph{explains} the passing of the test on \emph{all} good lines and points and not just a subset of them as given by the analogous interpolation steps in \cite{AroraS2003,HarshaKSS2024}. This argument for interpolation was also discovered in an independent work of Kominers, Thaler and Zheng \cite{KominersTZ2026} where the goal was to obtain improved soundness guarantees for the (non-derandomized) line-point test.

We now prove the following lemma.

\begin{lemma}\label{lem:existence-of-low-deg-Q}
Let $\F$ be any field of size $q$ and let $p$ denote its characteristic. 

There exists a non-zero polynomial $Q(X,Y,Z) \in \F[X,Y,Z]$ with $(1,1,\hat{k})$-degree $D$ at most $O\left( \frac{\hat{k}}{\epsilon^2} \right)$ such that for every \emph{good} line $L$, and the corresponding line polynomial $g_{L}$ of degree $k \leq \hat{k}$, the univariate polynomial $Q((X,Y)_L,g_{L})$ is identically zero and $Q(\va,f(\va))$ is zero for every \emph{good} point $\va$. 

\end{lemma}

We first set the parameters needed for the lemma. 

\paragraph{Setting parameters: } The parameter $D$ is set to be equal to $\left(\frac{12\cdot 3 \cdot 4 \cdot 400}{2}\frac{\hat{k}}{\epsilon^2} \right)$ and the parameter $\rho$ is set to $\frac{12\cdot 3 \cdot 1600\cdot 1600}{5}\cdot \frac{\hat{k}}{\epsilon^4 rq}$. Recall that in the context of the results in this paper, the parameter $\epsilon \in (0,1]$ is an arbitrary constant, $r = (1/\epsilon)^c$ for a large enough $c$, and $k \leq \hat{k} = o(q)$. We have the following simple claim summarizing some of the inequalities satisfied for our setting of parameters. 

\begin{claim}\label{clm:inequalities-in-parameters}
Let $\hat{k}, \rho, D$ be parameters as defined earlier in this section. Then, for all sufficiently large $q$,
\begin{itemize}
    \item $1 < \epsilon r/20 < \frac{1}{2\rho}$ 
    \item $D > 20\hat{k} \geq 20k$ 
    \item $\rho qr = \Theta(\hat{k})$
    \item $\exp(-\frac{\epsilon \rho qr}{100}) < 1/q^{100}$
\end{itemize}
\end{claim}

We are now ready to proceed with the proof. 

\begin{proof}[Proof of \cref{lem:existence-of-low-deg-Q}]
    The proof of existence of $Q$ will be via the probabilistic method. We will probabilistically pick a small subset $S$ of all good lines, and find a polynomial $Q$ with low $(1,1,k)$-degree such that for every line $L$ in $S$, $Q((X,Y)_L,g_{L})$ is identically zero via the polynomial method. We will then use the geometry of the set up to show that with high probability, any such $Q$ also satisfies the conclusion of the lemma. We now discuss the details. 

    \paragraph{Constructing $S$: } The set $S$ is constructed by starting with an empty set and including every good line $L$ in $S$ independently with probability $\rho$. Thus, the expected size of the set $S$ equals $\rho\cdot |\text{Good Lines}|$. We also know that the number of good lines is at least $\frac{\epsilon qr}{10} \geq \frac{4\epsilon qr}{5}$, and at most $ qr$. Thus, 
    \[
   \frac{\epsilon \rho qr}{10} \leq  \E[|S|] \leq \rho rq \, .
    \]
By Chernoff bound (\cref{lem:chernoff}), we also get that with probability at least $\left(1-\exp\left(-\frac{\epsilon \rho qr}{100}\right)\right)$, 
    \[
    \frac{1}{2}\E[|S|] \leq |S| \leq \frac{3}{2}\E[|S|] \, .
    \]
For the rest of the proof, we assume that the above tail bound holds. 

    \paragraph{Constructing $Q$: } For this step, we proceed via the polynomial method. Let $Q(X,Y,Z)$ be a polynomial with $(1,1,\hat{k})$-degree $D$ such that its coefficients are formal variables. 
    The goal is to show that there is a non-zero specialization of these formal variables (for coefficients) such that the $Q$ satisfies   $Q((X,Y)_L,g_{L}) \equiv 0$ for every $L$ in the set $S$. Let the line $L$ be defined as 
    \[
    L = \{(x_0,y_0) + \tau\cdot (u_0,u_1) : \tau \in \F\} \, .
    \]
    Thus, the univariate $Q((X,Y)_L,g_L)$ can be explicitly written as $R_L(T) := Q\left(x_0 + Tu_0, y_0 + Tu_1, g_L(T)\right)$. The coefficients of $R_L(T)$ are clearly homogeneous linear forms in the coefficients of $Q$, and the degree of $R_L$ is at most the $(1,1,\hat{k})$-degree of $Q$, which is at most $D$. Thus, setting $R_L(T)$ to be identically zero imposes at most $(D+1)$ homogeneous linear constraints on the coefficients of $Q$. If the number of such constraints is strictly less than the number of variables in the linear system, we must have a non-zero solution. Thus, from \cref{lem:monomial-count}, we have that there is a non-zero solution if
    \begin{align*}
        \frac{D^3}{12\hat{k}} \geq (D+1)|S|. 
    \end{align*}
    Since $D$ is a natural number, $D+1 \leq 2D$ and $|S|$ is at most $\frac{3}{2}\E[|S|]$, which in turn is at most $\frac{3}{2} \rho rq$, it suffices to have $D$ such that $D^2 > 36\hat{k}\rho q r$, which is an inequality satisfied by our choice of $D, \rho$. 
We note that this step is completely deterministic once the random bits in the choice of the set $S$ are fixed. 

    \paragraph{Behavior of $Q$ on other good lines:} We now argue that with high probability (over the choice of $S$), any polynomial $Q$ constructed in the previous step satisfies $Q((X,Y)_L, g_L) \equiv 0$ for every good line $L$.  

    Let $L$ be an arbitrary good line that is not in the set $S$. By definition, $L$ has at least $\epsilon q/10$ good points on it, and through each of these good points, there are at least $\epsilon r/10$ lines that are also good and consistent with the value of $f$ at the point. Let $\va$ be any good point on the line $L$ and let $R_L(T)$ be the polynomial $Q((X,Y)_L, g_L)$. The goal in this step is to show that the polynomial $R_L(T)$ is identically zero with high probability over the randomness in the choice of $S$. By a union bound over the set of all good lines, this will complete the proof of the lemma. To this end, we proceed by showing that $R_L(T)$ has \emph{too} many zeroes, or in other words, the polynomial $Q((X,Y)_L, g_L)$ vanishes at many points on $L$, and hence must be identically zero. 
    
    The key observation for this is that for any point $\va$, if $L$ and $L'$ are both neighbors of $\va$ in the graph $G'$, i.e. the test passes for both the line-point pairs $(L',\va)$ and $(L,\va)$,  then the line polynomials $g_L$ and $g_{L'}$ agree on $\va$. Moreover, if the line $L'$ is in $S$, then, by construction of $Q$, $Q((X,Y)_{L'}, g_{L'})$ is identically zero. From the above two properties, we have that $R_L$ also has a zero at $\va$. We state the following simple claims and use them to complete the proof of the lemma, but we need a definition first. 

    We say that a good point $\va$ on $L$ is \emph{fortunate} if there is a line $L' \neq L$ through $\va$ such that $L'$ is a neighbor of $v$ in the graph $G'$ and $L'$ is contained in $S$. A point that is not fortunate is said to be \emph{unfortunate}.   
    \begin{claim}\label{clm:Q-construction-1}
       $\va$ is unfortunate with probability at most $(1-\rho)^{\left(\frac{\epsilon r}{10}-1\right)}$, which, from our choice of parameters is at most $ \left( 1 - \frac{\epsilon \rho r}{40} \right)$.
    \end{claim}

    \begin{claim}\label{clm:Q-construction-2}
    Let $\va, \va'$ be two distinct good points that are neighbors of $L$ in the graph $G'$. Then, the events that $\va$ and $\va'$ are fortunate are independent of each other. 
    \end{claim}

    Since $L$ is a good line, it has at least $\epsilon q/10$ good points as neighbors in $G'$ and from \cref{clm:Q-construction-1}, we have that the expected number of fortunate points on the line $L$ is at least $ \frac{\epsilon^2 \rho q r}{400}$, which from the choice of our parameters is $\Theta(\hat{k}/\epsilon^2)$, and from our choice of constants, this is  at least $10D$. Furthermore, from \cref{clm:Q-construction-2}, the number of fortunate points on $L$ is a sum of independent random variables, and hence via \cref{lem:chernoff}, we get that with probability at least $\left(1 - \exp\left(-\Omega\left({\hat{k}/\epsilon^2} \right)\right)\right)$, $L$ contains at least $(D+1)$ fortunate points. This implies that the univariate polynomial $R_L = Q((X,Y)_L, g_L)$ that has degree at most $D$ has at least $D+1$ distinct zeroes, and hence must be identically zero. 
    
    From a union bound over the at most $qr < q^2$ good lines, we get that with high probability $Q$ satisfies $$Q((X,Y)_{\hat{L}}, g_{\hat{L}}) \equiv 0$$ for every good line $\hat{L}$. 
    
    We now note that the above condition also implies that $Q$ must satisfy $Q(\va, f(\va)) = 0$ for every good point $\va$, since every good point is incident to many good lines $L$ (all of its neighbors in the graph $G'$) such that for each of these lines $f$ and $g_L$ agree at $\va$. This condition lets us conclude that $\deg_Z(Q)$ is at least one: if this was not the case, $Q$ is a bivariate of total degree at most $D$ that vanishes on at least $4\epsilon/5 q^2$ points (namely, all the good points) in $\F^2$. But, $D = o(q)$, so this implies that $Q$ must be identically zero, which is a contradiction. Finally, since $Q$ depends on $Z$ and yet does not contain any monomial with its $Z$-degree divisible by $p$, we get from the definition of Hasse derivatives that $\partial_Z(Q)$ must be non-zero. 

    This completes the proof of \autoref{lem:existence-of-low-deg-Q}.

\end{proof}

\paragraph*{Proof of technical claims}
\begin{proof}[Proof of \cref{clm:Q-construction-1}]
Every good line incident to $\va$ is included in the set $S$ independently with probability $\rho$ and $\va$ being a good point is a neighbor of at least $\epsilon r/10$ such good lines in $G'$, including $L$ itself. So, the probability that none of the good lines incident to $v$, which are not equal to $L$, is contained in $S$ is at most $(1-\rho)^{\left(\frac{\epsilon r}{10}-1\right)}$. The rest of the argument is just a sequence of inequalities using the fact that  $1 < \epsilon r/20 < \frac{1}{2\rho}$.
\begin{align*}
(1-\rho)^{\left(\frac{\epsilon r}{10}-1\right)} &\leq (1-\rho)^{\left(\frac{\epsilon r}{20}\right)} \\
&\leq 1 - \rho\frac{\epsilon r}{20} + \frac{1}{2}\left(\rho \frac{\epsilon r}{20}\right)^2 \cdot \left(1 + \sum_{i=1}^{\infty} \left(\rho \frac{\epsilon r}{20}\right)^i\right)\\
&\leq 1 - \rho\frac{\epsilon r}{20} + \frac{1}{2}\left(\rho \frac{\epsilon r}{20}\right)^2 \left( 1 + \sum_{i = 1}^{\infty} 1/2^i\right) \\
& \leq 1 - \rho\frac{\epsilon r}{20} + \left(\rho \frac{\epsilon r}{20}\right)^2 \\
& \leq 1 - \rho\frac{\epsilon r}{40}
\end{align*}

\end{proof}

\begin{proof}[Proof of \cref{clm:Q-construction-2}]
    The independence of the events at $\va, \va'$ immediately follows from the fact that every line is included in $S$ independently with probability $\rho$ and the events at $\va, \va'$ do not have any line in common apart from $L$ itself, which does not play a role in these points being fortunate. 
\end{proof}

\subsection{Proof of \cref{thm:main technical}}
We are now ready to complete the analysis of the low-degree test.

\begin{proof}[Proof of \autoref{thm:main technical}]

    Since the function $f$ passes the test with probability $\epsilon$, we get from \autoref{lem:existence-of-low-deg-Q} that there is a $Q(X,Y,Z)$ with $(1,1,\hat{k})$-degree at most $O\left( \frac{\hat{k}}{\epsilon^2} \right)$ such that for every good line $L$, and the corresponding line polynomial $g_{L}$ of degree $k$, the univariate polynomial $Q((X,Y)_L,g_{L})$ is identically zero.  

    Note that $Q$ must depend on the variable $Z$, else, it has degree at most $D$ and vanishes at all good points, and there are much more than $Dq$ of them for our choice of parameters. Thus, if $Q$ does not depend on $Z$, then by \autoref{lem:SZ}, it must be identically zero.

    \paragraph*{Reducing to the non-degenerate case: }

    As in the proof of \autoref{thm:gen-PS-parallel-lines}, we work with the square-free part of $Q$. Let $Q_1$ be the square-free part of $Q$. Let $s \in \N$ be the largest integer such that there exists a polynomial $\hat{Q}(X,Y,Z)$ for which $Q_1(X,Y,Z) = \hat{Q}(X,Y,Z^{p^s})$.

This implies that $\partial_Z(\hat{Q})$ is a non-zero polynomial. We now note that $\hat{Q}$ also satisfies the interpolation constraints on all good lines and good points, i.e.  $\hat{Q}((X,Y)_L,\hat{g}_{L}) \equiv 0$ for all good lines $L$ and line polynomials $\hat{g}_L := g_L^{p^s}$ and $\hat{Q}(\va,\hat{f}(\va))$ is zero for all good points $\va$, where $\hat{f}(\va) := f(\va)^{p^s}$. Furthermore, the $(1,1,p^{s}\hat{k})$-degree of $\hat{Q}$ is at most $D$. Notice that $p^s < \deg_Z(Q_1)$. 

    Now, if $\deg_Z(\partial_Z(\hat{Q}))$ is zero, i.e. it is a non-zero polynomial of degree at most $D$ in $\F[X,Y]$, then, it follows from \autoref{lem:SZ} that there must exist a good point $\va$ such that $\partial_Z(\hat{Q})$ is non-zero for $\va$. We now argue that a similar conclusion holds even when $\deg_Z(\partial_Z(\hat{Q}))$ is at least one. To see this, we consider the discriminant of $\hat{Q}$ with respect to $Z$. From \autoref{thm:disc-and-sqr-free-multivariates}, and the properties of $\hat{Q}$ discussed so far, we get that $\disc_Z(\hat{Q})$ is a non-zero bivariate of degree at most $O(\deg_Z(Q)D)$. Multiplying $\disc_Z(\hat{Q})$ with the leading coefficients of $\hat{Q}$ and $ \partial_Z(\hat{Q})$, while viewing these as a univariate in $Z$, we get a non-zero polynomial of degree at most $O(\deg_Z(\hat{Q})D)$. Since $q$ is large enough compared to $\poly(\deg_Z(Q)D\hat{k})$, we get from \autoref{lem:SZ} that there is a good point $\va$  for which this product polynomial is non-zero. From \autoref{obs:resultant-restrictions-commute} and \autoref{thm:disc-and-sqr-free}, we get that for any such $\va$, the univariates $\hat{Q}(\va, Z)$ and $\partial_Z(\hat{Q})(\va,Z)$ do not have a common-zero. In particular, since $\hat{f}(\va)$ is a zero of  $\hat{Q}(\va, Z)$ (from the interpolation constraints), we get that $\partial_Z(\hat{Q})(\va,\hat{f}(\va))$ must be non-zero. For the rest of the proof, we work with this good point $\va$.

  \paragraph*{Setting up for lifting from parallel lines: }  
     We now proceed with the proof of the theorem. Let $\hat{\calL}$ be the neighbors of the point $\va$ in the graph $G'$. Since $\partial_Z(\hat{Q}(\va,\hat{f}(\va))) \neq 0$ and for every $L \in \hat{\calL}$, $\hat{g}_L$ and $\hat{f}$ must agree on $\va$, we get that the polynomial $\partial_Z(\hat{Q}((X,Y)_L,Z))$ is non-zero at $Z = \hat{g}_L$. Let us fix one such line $L$ in $\hat{\calL}$, and consider the set ${\calL '}$ of lines $L'$ such that $L$ and $L'$ have a common neighbor in $G'$. From the properties of $G'$, $L$ is adjacent to $\epsilon q/10$ points and each of these points is incident to at least $\epsilon r/10$ lines. Thus, $|{\calL '}|$ is at least $\epsilon^2 qr/100$. Since there are at most $q$ lines in any direction, we get that there is a set $S \subseteq [r]$ of size at  least $\epsilon^2 r/200$, such that for every $i \in S$, $|\calL_i \cap \calL'|$ is at least $\epsilon^2 q/200$. We focus on these directions. 

    We now note that for every $i \in S$, the set of parallel lines $\calL_i \cap \calL'$ together with the line $L$ satisfies the hypothesis of \autoref{lem:lifting-parallel-lines-general}, with the degree of the root on the line polynomials being at most $k' = p^s k$. Thus, it follows that if $\epsilon^2 q/200 > \Omega(\deg_Z(Q)Dk')$, (which is an inequality that holds in our setting) then for each such $i$, we have a rational function $U_i(X,Y)$ of degree at most $Dk' = p^sDk$ that is a $Z$-root of $\hat{Q}(X,Y,Z)$ and furthermore, the restriction of $U_i$ on $L$ equals $\hat{g}_L$ for all but $O(Dk')$ lines in $\calL_i \cap \calL'$. 

\paragraph*{From rational roots to polynomial roots: }
    In our setting of parameters,   $\epsilon^2 r/200$ is greater than the $Z$-degree of $\hat{Q}$ (if the constant $c$ in \autoref{thm:main technical} is chosen to be large enough). So, there must exist distinct $i,j \in S$ such that $U_i(X,Y) = U_j(X,Y)$. This rational function, that we now refer to as $U$, has degree at most $Dk'$ and for at least $\epsilon^2 q/200 - O(Dk') = \Omega(\poly(\epsilon) q)$ many parallel lines in two different directions, the restriction of $U$ to these lines is a polynomial of degree at most ${kp^s}$. In other words, for each of these lines, the restriction of the denominator of $U$ to this line divides the restriction of the numerator of $U$ on this line, and yields a quotient of degree at most $kp^s$. Thus, from \autoref{lem:PS}, we get that if $q > \poly(1/\epsilon) D\hat{k}$, then when reduced to its simplest form, the rational function $U$ is actually a polynomial of degree at most $2kp^s$. Moreover, the restriction of $U$ on at least $\poly(\epsilon)(q-D\hat{k}) = \poly(\epsilon)q$ of these lines $L \in \calL_i \cup \calL_j$ equals the corresponding line polynomial $\hat{g}_L$. Since each of these lines is incident to at least $\Omega(\epsilon q)$ points in the graph $G'$ and the line polynomials $\hat{g}_L$ are consistent with the function values of $\hat{f}$ at these points, we get that $U$ equals the function $\hat{f}$ on $\poly(\epsilon) q^2$ points.  

    We also know that the polynomial $U$ is a $p^s$ power when restricted to $\poly(\epsilon)q$ lines in two different directions. Thus, by (an iterative application of) \autoref{lem:pth-power-on-lines}, we get that there exists a polynomial $\tilde{U}(X,Y)$ of degree at most $2k$ such that $U := \tilde{(U)}^{p^s}$. Furthermore, since $p$th roots of a field element in $\F$ are unique, and $\hat{f}(\va) = f(\va)^{p^s}$ by definition, we get that if $U$ agrees with $\hat{f}$ on the point $\va$, then $\tilde{U}$ must agree with $f$ on $\va$. Furthermore, the degree of $\tilde{U}$ is at most $\deg(U)/p^s \leq 2k$.

    This completes the proof of the theorem. 
\end{proof}

\section{Counterexamples when $k \gg q^{1/2}$}\label{sec:counterexamples}

In this section, we show some limitations of  derandomized tests when $q \leq (k/\eps)^2$. The main result of this section is the following theorem which immediately implies \cref{thm:intro-counterexample}.

\begin{theorem}\label{thm:counterexamples}
    For every $r, s \in N$ and prime $p$, the following holds.
        
    Let the local agreement parameter $\eps := p^{-r}$. Let $\F= \F_q$ be the finite field of size $q := p^{2rs}$ and the degree parameter $k := p^{r(s-1)} = \eps\sqrt{q}$. 
    
    Let $\calL$ be the set of all lines in $\F_q^2$
    whose slope lies in the subfield $\F_{p^{rs}}$. Then there exists a collection of functions $\left(f_\ell:\ell \to \F_q\right)_{\ell \in \calL}$ such that:
    \begin{enumerate}
        \item \label{item:flowdegree} For each $\ell \in \calL$, $f_\ell$ agrees with a univariate polynomial $P_\ell$ of degree at most $k$ on $\ell$.
        \item \label{item:twolinesagree} If we pick the two lines $\ell, \ell' \in \calL$ independently and uniformly at random from the set of all lines in $\calL$, and let $\va$ be their point of intersection
        (undefined if the lines are parallel), we have:
        \[ \Pr[ f_\ell(\va) = f_{\ell'}(\va) ] \geq \eps - O\left(\frac{1}{\sqrt{q}}\right)\,.\]
        \item \label{item:psfails} There exists a trivariate polynomial $Q(X,Y,Z)$ of $Z$-degree
        $1/\eps$ and $(1,1,k)$-degree at most $\sqrt{q}+1$ such that for each
        $\ell \in \calL$, $P_\ell$ is a $Z$-root of the restriction of $Q$ to
        $\ell$, yet $Q$ has no polynomial $Z$-root.
        \item \label{item:noglobalagreeforlines}For any bivariate polynomial $g(X,Y)$ of total degree at most $Ck$ for some $C<\sqrt{q}$, if we choose a line $\ell \in \calL$ uniformly at random, then we have:
        \begin{align*}
        \Pr [ g|_{\ell} = f_\ell ] &\leq \nicefrac{C}q\\
        \E [ \agree( g|_{\ell}, f_\ell) ] &\leq \nicefrac{2C\eps}{\sqrt{q}}\,.
         \end{align*}
        \item \label{item:derandldtfails} There exists a function $f\colon \F_q^2 \to \F_q$ satisfying
        \begin{align*}
        \E_{\ell \in \calL}[\agr(f|_\ell,f_\ell)] &\geq \eps,
        \end{align*}
        and yet $\agr(f,g)\leq \nicefrac{C}{\sqrt{q}}$ for every bivariate polynomial $g(X,Y)$ of total degree at most $C\sqrt{q}/\eps$-degree where $C>1$. 
    \end{enumerate}
\end{theorem}

First for some notation and helper claims that will help us prove this theorem. Given a prime $p$ and $r, s \in \N$, consider the field extensions: $E := \F_{p^r} \subseteq H := \F_{p^{rs}} \subseteq \F := \F_{p^{2rs}}$. It will be convenient to use the following notation: $e := p^r, h := |H|=e^s$ and $q := |\F|=e^{2s}=h^2$. To understand the counterexample, it is instructive to think of the case when $p =2$ and $r =1$. 

We will use $\Tr$ to denote the usual trace polynomial from $\F=\F_q$ to $E=\F_e$,
\[ \Tr(z) = \Tr_{\F/E}(z)=\sum_{i=0}^{2s-1} z^{e^{i}}\,.\]

Our construction will also use the {\em half-trace} function, denoted by $\HTr$, which refers to trace polynomial from $H=\F_h$ to $E=\F_e$, but extended as a polynomial to all of $\F$.
\[ \HTr(z) := \Tr_{H/E}(z)=\sum_{i=0}^{s-1} z^{e^{i}}\,.\]
Evaluated on $\F$ it is an $E$-linear map, 
and evaluated on $H$ it is the field trace map to $E$.

For $a \in H$ and $b \in \F$,
let $\ell_{a,b}\colon \F_q \to \F_q^2$ denote the line
\[ \ell_{a,b}(T)  := (T, aT + b)\,,\]
Note that
\begin{equation}\label{eq:call}
    \calL = \{ \ell_{a,b} \mid (a,b) \in H\times \F \}\,.
\end{equation}
For each $a \in H$ and $b \in \F$, define the polynomial $P_{a,b}(T) \in \F[T]$ of degree $e^{s-1}$ by
\[ P_{a,b}(T) := \HTr(b^{h} T)\,.\]
For the line $\ell = \ell_{a,b}$, we use $P_{a,b}$ to define the function $f_\ell\colon \ell \to \F$ as follows:
\begin{equation}\label{eq:linefunction}
f_\ell( t, at+b ) =  P_{a,b}(t)\,.
\end{equation}
First, we see that these line polynomials are mostly consistent. The lines $\ell_{a,b}$ and $\ell_{a',b'}$ intersect at $\left(\nicefrac{(b'-b)}{(a-a')}, \nicefrac{(ab'-a'b)}{(a-a')}\right)$. 
\begin{claim}\label{clm:consistency-of-line-poly}    
    \[ \Pr_{\substack{a, a' \in H \\ b,b'\in \F}} \left[ P_{a,b}\left(  \frac{b'-b}{a-a'} \right) = P_{a',b'}\left(  \frac{b'-b}{a-a'} \right) \right] \geq \frac{1}{e} - O\left(\frac{1}{h}\right)\,.\]
\end{claim}
\begin{proof}
    The relevant expressions are only defined if $a \neq a'$. Condition on $a \neq a'$; this only changes the claimed probability by $O(1/h)$.
    
    Let $t$ be the random variable $\frac{b'-b}{a-a'} \in \F_p$.
    We have:
    \begin{align*}
        P_{a,b}(t) - P_{a',b'}(t) &= \HTr(b^h t) - \HTr((b')^h t)\\
        &=\HTr( (b^h - (b')^h) t)\\
        &= \HTr \left(   (b -b')^h  \cdot \frac{b-b'}{a'-a} \right)\\
        &= \HTr \left( \frac{ (b-b')^{h+1}}{a'-a} \right)\,.
    \end{align*} 
    Now observe that when $b,b'$ are picked uniformly from $\F$, 
    then $b-b'$ is a uniform element of $\F$, and $(b-b')^{h+1} = \mathrm{Norm}_{\F/H}(b-b')$
    is essentially  (upto error $\frac{1}{q}$) a uniform nonzero element of $H$. We also have that $a'-a$ is essentially (upto error $\frac{1}{h}$) a uniform element of $H$.

    Thus the probability in question is, upto error $O(\frac{1}{h})$ the probability that $\HTr$ of a uniform 
    nonzero element $c$ of $H$ has  $\HTr(c) = 0$. 
    But for $c \in H$, we have that $\HTr(c) = \Tr_{H/E}(c)$, and the probability 
    that $\HTr(c)$ equals $0$ equals $1/e$.
    
    Thus the probability in question is at least  $1/e - O(1/h)$.
\end{proof}

Next we see that there is no bivariate polynomial $U(X,Y)$ that has high agreement with these line polynomials in a very strong sense.

\begin{claim}\label{clm:agreement-bivariate}
    For any $C<h$ and for all $U(X,Y)$ of degree at most $Ch/e$,
    \[
    \Pr_{\ell \in \calL} [U|_\ell = f_\ell] \leq \frac{C}{q}\,.\] 
\end{claim}

\begin{proof}
Let $B, Q$ be polynomials defined as follows. 
\begin{align*}
B(X,Y) &:= X Y^h  - X^h Y,\\
Q(X,Y,Z) &:= Z^e - Z + B(X,Y)\,.
\end{align*}
We would need the following two technical claims for the proof. 
\begin{subclaim}\label{clm:no-bivariate-root}
For any polynomial $U(X,Y)$, the polynomial $H(X,Y) := Q(X,Y,U(X,Y))$ is not identically zero. 
\end{subclaim}
\begin{subclaim}\label{clm:line-poly-are-roots}
    For every $a \in H, b \in \F$, the polynomial $P_{a,b}(T)$ is a $Z$-root of $Q(\ell_{a,b}(T), Z)$. 
\end{subclaim}
Given the sub claims, the proof of the claim is fairly straightforward. Let $U$ is a bivariate polynomial of degree $d\leq Ch$. Define 
\[ R(X,Y) := Q(X,Y,U(X,Y))\,.\]
 By \cref{clm:no-bivariate-root}, $R$ is a non-zero bivariate polynomial of degree at most $\max(ed, h+1)$ and its individual degree in both $X,Y$ is upper bounded by $\max(ed,h)$, where $d \leq Ch/e$ is the total degree of $U$. Since $Q(\ell_{a,b}(T), P_{a,b}(T))$ is identically zero for every $a \in H, b \in \F$, and $f_\ell(t) = P_{a,b}(t)$ for every line $\ell$ parameterized as $(t,at+b)$, we get that $U|_\ell = f_\ell$ implies that $R|_\ell(T) = Q(T,aT+b, U(T,aT+b))$ is identically zero. Here we use that fact that $\deg_T(R|_\ell) \leq \max(ed,h+1)\leq \max(Ch,h+1) \leq h^2$ since $C< h$. So, the equality of $U_\ell$ and $f_\ell$ as functions on $\ell$ implies that they are equal as formal polynomials and since $f_\ell = P_{a,b}$ is a formal $Z$ root of $Q((X,Y)_\ell,Z)$, so must $U_\ell$. Thus, every such line $\ell$ for which $U|_\ell=f_\ell$ gives us a \emph{distinct} factor of $R(X,Y)$ of the form $(Y-aX-b)$. So, the number of such lines is at most the $Y$ degree of $R$, which is at most $\max(ed,h)\leq Ch=C\sqrt{q}$. Finally, observing that $\calL$ has exactly $hq = q^{3/2}$ lines completes the proof of the claim. 
\end{proof}
We now prove the subclaims. 
\begin{proof}[Proof of \cref{clm:no-bivariate-root}]
    Consider the infinite power series $\psi := B + B^e + B^{e^2} + B^{e^3} + \cdots + \infty$. Observe that $Q(X,Y,\alpha+\psi)$ is identically zero for every $\alpha \in E$. Thus, $\{a+\psi \mid a \in E\}$ are $Z$-roots of $Q$ in the ring of power formal series in $X,Y$ over $\F$. Furthermore, observe that these roots are truly power series (and cannot be polynomials) since $B$ is homogeneous and each term $B^{e^i}$ contributes non-zero monomials of degree exactly $e^i(h+1)$ that can never be canceled out. Since the $Z$-degree of $Q$ equals $e=|E|$, it cannot have any more roots in this ring (and hence the polynomial ring). 
\end{proof}

\begin{proof}[Proof of \cref{clm:line-poly-are-roots}]
    Follows immediately by plugging in and checking that $Q(\ell_{a,b}(T),P_{a,b}(T))$ is identically zero for all $a \in H$ and $b\in \F$. 
    \begin{align*}
    Q(\ell_{a,b}(T),P_{a,b}(T)) &= \left(P_{a,b}(T)\right)^e - P_{a,b}(T) + T(aT+b)^h - T^h(aT+b)\\
    &= \left(\HTr(b^hT)\right)^e - \HTr(b^hT) + T^{h+1}(a-a^h) + (Tb^h-T^hb)\\
    &= \left(\sum_{i=0}^{s-1} (b^hT)^{e^i}\right)^e - \left(\sum_{i=0}^{s-1} (b^hT)^{e^i}\right) + (Tb^h-T^hb) && \text{[since $a^h=h$ as $a \in H$]}\\
    &= \left(\sum_{i=1}^{s} (b^hT)^{e^i}\right) - \left(\sum_{i=0}^{s-1} (b^hT)^{e^i}\right) + (Tb^h-T^hb)\\
    &=(b^hT)^{e^s} - (b^hT) + (Tb^h-T^hb)\\
    &=b^{h^2}T^h - b^hT + (Tb^h-T^hb)&& \text{[since $e^s=h$]}\\
    &=T^h(b^{h^2}-b)=0 && \text{[since $b^{q}=b$ as $b \in \F_{h^2}=\F_q$]}\\
    \end{align*}
    Note that we crucially use that $a^h = a$ and $b^q = b$. 
\end{proof}

We are now ready to prove \cref{thm:counterexamples}.
\begin{proof}[Proof of \cref{thm:counterexamples}]
    Let $p$ be a prime and $r, s \in \N$. Consider the field extensions: $E := \F_{p^r} \subseteq H := \F_{p^{rs}} \subseteq \F := \F_{p^{2rs}}$. Note, the local agreement parameter $\eps = p^{-r} = 1/e$ and the degree parameter $k = \eps\sqrt{q} = h/e= p^{r(s-1)}$.

    The set $\calL$ of all lines whose slopes lie in the subfield $H=\F_{p^{rs}}$ is given by \eqref{eq:call} and the collection of functions $\left(f_\ell\right)_{\ell \in \calL}$ be as in \eqref{eq:linefunction}.
    
    \cref{item:flowdegree} follows from the definition of polynomial $P_{a,b}$ which has degree at most $e^{s-1}=p^{r(s-1)}$ while \cref{item:twolinesagree} follows from \cref{clm:consistency-of-line-poly} and \cref{item:psfails} follows from \cref{clm:no-bivariate-root} and \cref{clm:line-poly-are-roots}.

   We now use \autoref{clm:agreement-bivariate} to prove \cref{item:noglobalagreeforlines}. Let $g$ be a bivariate polynomial of total degree at most $Ck = Ch/e$ for some $C<h$. By \cref{clm:agreement-bivariate}, we have the first subitem that $\Pr_{L \in \calL}[g|_\ell = f_\ell] \leq \nicefrac{C}q$. Now, to prove the second subitem, we note that $g|_\ell, f_\ell$ are both univariates polynomial on $\ell$ of degree at most $Ck$. Thus, if they are not equal, then they agree on at most $Ck/q$ points on $\ell$. So, using \cref{clm:agreement-bivariate}, we get 
    \begin{align*}
        \E_{\ell \in \calL}[ \agree( g|_{\ell}, f_\ell)] &\leq \Pr_{L \in \calL}[g|_\ell = f_\ell] \cdot 1 + \Pr_{\ell \in \calL}[g|_\ell \neq f_\ell] \cdot \frac{Ck}{q} \\
        &\leq \frac{C}q + \left(1-\frac{C}q\right)\frac{Ch}{eq} \leq \frac{C}q + \frac{C}{e\sqrt{q}}\\
        &\leq \frac{2C\eps}{\sqrt{q}}. 
    \end{align*}

    Finally, we are left with proving \cref{item:derandldtfails}. We now define the points table $f\colon \F_q^2 \to \F_q$ as follows. For every point $\va \in \F^2$, let $\calL_{\va}$ be the subset of $|H|=\sqrt{q}$ lines in $\calL$ that are incident to $\va$. From \autoref{clm:line-poly-are-roots}, and the form of the polynomial $Q$ there,  we get that for every $\va \in \F^2$ and $\ell \in \calL_{\va}$, $Q(\va, Z)$ has $Z$-degree exactly $e=1/\eps$ and $f_\ell$ is a $Z$-root of $Q((X,Y)_\ell, Z)$. Thus, for every $\ell \in \calL_{\va}$, $f_\ell(\va)$ equals one of the $e$ roots of $Q(\va, Z)$. We set $f(\va)$ to be the most popular value taken by $h_{\ell}(\va)$ as we range over the $|H|$ lines $\ell \in \calL_{\va}$. 

    We now argue that this $f$ has at least $\eps$-expected local agreement with polynomials of degree at most $k$.     
    Now, for every $\va$, we have that $\Pr_{\ell \in \calL_{\va}}[f(\va) = h_{\ell}(\va)]$ is at least $1/e=\eps$. Hence, 
    \begin{align*}
        \E_{\ell \in \calL}[\agr(f|_\ell,f_\ell)] &= \Pr_{\ell \in \calL, \va \in \ell}[f(\va) = f_\ell(\va)]\\
        &= \Pr_{\va \in \F^2, \ell \in \calL_\va}[f(\va) = f_\ell(\va)]\\
         &= \frac{1}{q^2} \sum_{\va \in \F^2}\Pr_{\ell \in \calL_{\va}}[f(\va)= f_{\ell}(\va)] \\
        &\geq \frac{1}{q^2} \sum_{\va \in \F^2} \eps  \\
        &\geq \eps 
    \end{align*}
    
    Finally, we have to argue that $f$ has low agreement with every $g(X,Y) \in \F[X,Y]$ of degree at most $O(\sqrt{q})$. To see this, recall from the definition of $f$ that for every $\va \in \F^2$, $Q(\va, f(\va)) = 0$. Furthermore, from \autoref{clm:no-bivariate-root} that for every \emph{polynomial} $g(X,Y) \in \F[X,Y]$, $Q(X,Y,g(X,Y))$ is a non-zero polynomial, and its degree is at most $\max(\sqrt{q}+1, e\cdot \deg(g))$. Thus, if $f(\va)$ equals $g(\va)$ for some $\va \in \F^2$, then $Q(X,Y,g(X,Y))$ must vanish at $\va$. But from \autoref{lem:SZ}, this can happen on at most $\frac{\max(\sqrt{q}+1, e\cdot \deg(g))}{q}$ fraction of points in $\F^2$. If $\deg(g) \leq C\sqrt{q}/\eps$ for some $C \geq 1$, then this bound is at most $C/\sqrt{q}$. 
    
    This completes the proof of \cref{thm:counterexamples}. 
\end{proof}

{\let\thefootnote\relax
\footnotetext{\textcolor{\gitinfonotecolour}{\gitinfonote \easteregg}
}} 
{\small 
  \bibliographystyle{prahladhurl}
  \bibliography{derandLDT-bib.bib}
}
\end{document}

%% file: thmmacros.tex
\usepackage{amsthm}
\usepackage{thmtools,thm-restate}
\usepackage[nameinlink,capitalise]{cleveref}

\numberwithin{equation}{section}
\declaretheoremstyle[bodyfont=\it,qed=\qedsymbol]{noproofstyle}

\declaretheorem[numberlike=equation]{observation}

\declaretheorem[name=Observation,numbered=no]{observation*}

\declaretheorem[numberlike=equation]{subclaim}

\declaretheorem[numberlike=equation]{theorem}

\declaretheorem[name=Theorem,numbered=no]{theorem*}

\declaretheorem[numberlike=equation]{lemma}
\declaretheorem[name=Lemma,numbered=no]{lemma*}

\declaretheorem[numberlike=equation]{corollary}
\declaretheorem[name=Corollary,numbered=no]{corollary*}

\declaretheorem[name=Proposition,numbered=no]{proposition*}

\declaretheorem[numberlike=equation,name=Claim]{claim}
\declaretheorem[name=Claim,numbered=no]{claim*}

\declaretheorem[name=Conjecture,numbered=no]{conjecture*}

\declaretheorem[name=Question,numbered=no]{question*}

\declaretheoremstyle[bodyfont=\it,qed=$\lozenge$]{defstyle} 

\declaretheorem[numberlike=equation,style=defstyle]{definition}
\declaretheorem[unnumbered,name=Definition,style=defstyle]{definition*}

\declaretheorem[unnumbered,name=Example,style=defstyle]{example*}

\declaretheorem[unnumbered,name=Notation=defstyle]{notation*}

\declaretheorem[unnumbered,name=Construction,style=defstyle]{construction*}

\declaretheorem[unnumbered,name=Remark,style=defstyle]{remark*}


%% file: bibmacros.tex
\usepackage{nth}
\usepackage{intcalc}
\usepackage{etoolbox}
\usepackage{xstring}

\usepackage{ifpdf}
\ifpdf
\else
\usepackage[quadpoints=false]{hypdvips}
\fi

\newcommand{\ECCCurl}[2]{http://eccc.hpi-web.de/report/\ifnumcomp{#1}{>}{93}{19}{20}#1/#2/}

\newcommand{\shortECCC}[2]{\texttt{\href{http://eccc.hpi-web.de/report/\ifnumcomp{#1}{>}{93}{19}{20}#1/#2/}{eccc:TR#1-#2}}}

\newcommand{\parseECCC}[1]{
\StrSubstitute{#1}{TR}{}[\tmpstring]%
\IfSubStr{\tmpstring}{/}{ 
\StrBefore{\tmpstring}{/}[\ecccyear]%
\StrBehind{\tmpstring}{/}[\ecccreport]%
}{
\StrBefore{\tmpstring}{-}[\ecccyear]%
\StrBehind{\tmpstring}{-}[\ecccreport]%
}%
\shortECCC{\ecccyear}{\ecccreport}}